\documentclass[preprint,onecolumn,nofootinbib]{revtex4}
\usepackage[colorlinks=true,linkcolor=blue,urlcolor=blue,filecolor=black,citecolor=red,pdfstartview=FitV,pdftitle={},pdfsubject={},pdfkeywords={},pdfpagemode=None,bookmarksopen=true]{hyperref}
\usepackage{graphicx}
\usepackage{amsmath}
\usepackage{amsfonts}
\usepackage{amssymb,ulem}
\usepackage{color}%
\usepackage{tikz}
\usepackage{dcolumn}
\usepackage{xcolor}

\begin{document}
\title{
Mixed-state entanglement and phase transitions in Einstein-Born-Infeld massive gravity
}
\author{Zhe Yang $^{1,2}$}
\email{yzar55@stu2021.jnu.edu.cn}
\author{Jian-Pin Wu $^{1}$}
\email{jianpinwu@yzu.edu.cn}
\thanks{Corresponding author}
\author{Peng Liu $^{2}$}
\email{phylp@email.jnu.edu.cn}
\thanks{Corresponding author}
\affiliation{
  $^1$ Center for Gravitation and Cosmology, College of Physical Science and Technology, Yangzhou University, Yangzhou 225009, China\\
  $^2$ Department of Physics and Siyuan Laboratory, Jinan University, Guangzhou 510632, China\\
}

\begin{abstract}

 We study mixed-state entanglement measures in Einstein-Born-Infeld (EBI) massive gravity theory, a model exhibiting both Hawking-Page phase transitions and effective metal-insulator transitions (MIT) at finite temperatures. Our comprehensive investigation reveals that the entanglement wedge cross-section (EWCS), a holographic probe of mixed-state entanglement structure, demonstrates distinctive properties in detecting phase transitions. For the effective MIT, we find that the higher-order terms of EWCS align closely with the crossover temperature, a feature not shared by holographic entanglement entropy (HEE). This sensitivity makes EWCS a complementary geometric tool for probing effective phase transitions at finite temperatures. In Hawking-Page phase transitions, we observe that all entanglement measures effectively diagnose both first-order and second-order phase transitions, with EWCS showing configuration-independent behavior. Importantly, we discover that all geometry-related quantities, including entanglement measures, demonstrate a universal critical exponent of 1/3 near the second-order phase transition point. This result suggests a fundamental connection between quantum information theory and critical phenomena in gravitational systems, and also highlights the potential of EWCS as a powerful probe for phase transitions.

\end{abstract}

\maketitle
\tableofcontents

\section{Introduction}
\label{sec:introduction}
Quantum entanglement is one of the most fundamental properties of quantum systems, playing an important role in condensed matter theory, quantum information, and holographic gravity. Recent research has demonstrated that quantum information can detect quantum phase transitions and play a key role in spacetime emergence \cite{Eisert2006EntanglementIQ, Osterloh:2002na, Amico:2007ag, Wen:2006topo, Kitaev:2006topo}. Over the years, various entanglement measures have been proposed, including entanglement entropy (EE), mutual information (MI), and R\'enyi entropy. However, EE is not suitable for characterizing the more prevalent mixed-state entanglement. To address this limitation, several mixed-state entanglement measures have been developed, such as entanglement of purification, reflected entropy, and quantum discord \cite{Vidal:2002zz,plenio2005logarithmic, Horodecki:2009zz}. Nevertheless, calculating mixed-state entanglement measures remains challenging, particularly in strongly correlated systems.

Holographic duality has proven to be a powerful tool for studying strongly correlated systems. Within this framework, a strongly correlated system is dual to a classical gravitational system \cite{tHooft:1993dmi, Susskind:1994vu, Maldacena:1997re, Witten:1998qj, Hartnoll:2016apf}. Many investigations have revealed that quantum information is dual to the background geometry in the gravitational theory. For instance, entanglement entropy (EE) is one of the most widely used quantum information measures. The holographic dual of EE corresponds to the minimal surface in the bulk, known as holographic entanglement entropy (HEE) \cite{Ryu:2006bv}. HEE has been extensively employed to study thermal phase transitions in holographic gravity theories \cite{Caceres:2015vsa, Zeng:2015tfj, Cai:2012nm, Peng:2014ira, Peng:2015yaa}. However, studies have revealed that HEE is susceptible to thermal entropy contributions in mixed-state systems \cite{Ling:2015dma, Yang:2023wuw}. Similarly, the holographic realization of MI, referred to as holographic mutual information (HMI), is also affected by thermal entropy, since it is constructed from HEE. To probe mixed-state entanglement through an independent geometric structure, the entanglement wedge cross-section (EWCS) has been proposed as a novel holographic probe of mixed-state entanglement. Recent studies indicate that EWCS may correspond to various boundary quantum information quantities, including entanglement of purification, reflected entropy, logarithmic negativity, and odd entropy \cite{Takayanagi:2017knl,Kudler-Flam:2018qjo, Dutta:2019gen, Jokela:2019ebz, BabaeiVelni:2019pkw, Vasli:2022kfu, Camargo:2022mme, Tamaoka:2018ned, Nguyen:2017yqw}. While the precise boundary dual of EWCS is still an open question, it is well established that EWCS provides a robust probe of mixed-state entanglement in strongly coupled systems \cite{Liu:2019qje, Huang:2019zph, Liu:2020blk, Chen:2021bjt, Li:2021rff, Chowdhury:2021idy, Sahraei:2021wqn, ChowdhuryRoy:2022dgo, Maulik:2022hty, Jain:2020rbb, Jain:2022hxl, Yang:2023wuw}. In what follows, we therefore refer to EWCS simply as a mixed-state entanglement probe, without committing to a specific boundary dual. This allows us to compare it with HEE and HMI on equal footing, since all three quantities probe entanglement through different geometric structures in the bulk.

Born-Infeld (BI) theory represents a distinctive class of nonlinear electromagnetic theory, initially developed to address the point charge singularity in classical electromagnetic field theory \cite{born1934foundations}. Hoffmann pioneered the investigation of Einstein gravity coupled with BI electrodynamics \cite{hoffmann1935gravitational}. Subsequently, it was discovered that BI-type effective actions naturally emerge in the context of open superstring theory and D-brane dynamics \cite{wiltshire1988black, cataldo1999three, Ferrara:2015ixa}. This connection has significantly broadened the relevance of BI theory beyond its original electromagnetic formulation, establishing it as a crucial element in modern theoretical physics.
In the last two decades, numerous studies have explored BI theory as a gravitational theory and investigated the properties of BI black holes \cite{Breton:2003tk, Hendi:2012zz, Wang:2019kxp}. BI-type actions arise in holographic constructions that model aspects of strongly coupled systems, including quantum liquids, and certain condensed matter systems with novel transport properties \cite{Kundu:2013eba, Karch:2009zz, Baggioli:2016oju, Kiritsis:2016cpm, Cremonini:2017qwq}. Additionally, massive gravity theory, known for its capacity to break translational symmetry, plays a crucial role in inducing momentum dissipation within holographic duality. This feature corresponds to systems exhibiting finite DC conductivity in condensed matter theory \cite{deRham:2014zqa, Zeng:2014uoa, Vegh:2013sk}. Several studies have demonstrated that an effective metal-insulator transition (MIT) can occur in this model, with transport properties changing as temperature varies \cite{Baggioli:2014roa, Baggioli:2016oju}. It should be noted that an effective MIT refers to a crossover in transport, typically identified by a pronounced suppression of the DC conductivity $\sigma_{\mathrm{DC}}$, and a change in the sign of $\sigma_{\mathrm{DC}}'(T)$. This behavior can occur without an abrupt change in the ground state or the emergence of a well-defined order parameter. By contrast, a quantum phase transition takes place strictly at $T\to 0$ and is characterized by a critical point that separates distinct phases. In practice, the microscopic origins of effective MIT are generally different from those of a true quantum phase transition. Beyond transport properties, holographic quantum information in massive gravity has also been studied in the context of thermodynamic phase transitions \cite{Cai:2014znn, Zeng:2015tfj, Liu:2021rks}.
The effective MIT is analyzed in detail in Sec.~\ref{sec:mit} using holographic mixed-state entanglement measures.

These findings motivate the combination of these two theories to construct a nonlinear electromagnetic field coupled with massive gravity, known as Einstein-Born-Infeld (EBI) massive gravity theory \cite{Hendi:2015hoa}. The EBI massive gravity theory, characterized by its nonlinear electrodynamics field, may exhibit novel transport properties. However, the mixed-state entanglement measures in this theory and their relationship to transport properties remain unexplored. This model can exhibit two distinct types of phase transitions: the effective metal-insulator transition, which occurs as temperature increases, and the Hawking-Page phase transition. These phenomena provide valuable insights into the thermodynamic behavior and critical properties of the system across different regimes. Therefore, our study aims to systematically examine mixed-state entanglement measures during these two phase transitions and characterize their behavior.

This paper is organized as follows. In Sec.~\ref{enbi}, we present the holographic setup of EBI massive gravity theory and introduce the holographic entanglement measures. In Sec.~\ref{sec:mit}, we investigate the effective MIT and explore the relationship between this phase transition and holographic entanglement measures. In Sec.~\ref{sec:hawkingpage}, we discuss the correlation between the Hawking-Page phase transition and various holographic quantum information quantities, including HEE, HMI, and EWCS. We further examine the scaling behavior of different entanglement measures in Sec.~\ref{sec:sca}. Finally, we summarize our findings in Sec.~\ref{sec:discuss}.

\section{Holographic setup for EBI massive Gravity theory and mixed-state entanglement measures}
\label{enbi}

We begin by discussing the model of massive gravity with a nonlinear electrodynamics field, which is called EBI massive gravity theory. Following that, we introduce the mixed-state entanglement measures, including HEE, HMI, and EWCS.

\subsection{Einstein-Born-Infeld massive gravity theory}
The d-dimensional action of EBI massive gravity system reads \cite{Hendi:2015hoa},
\begin{equation} \label{eq1}
  \mathcal{L}=-\frac{1}{16\pi}\int d^dx\sqrt{-g}\left[\mathcal{R}-2\Lambda+4\beta^2\left(1-\sqrt{1+\frac{\mathcal{F}}{2\beta^2}}\right)+m^2\sum\limits_{i}^4c_i\mathcal{U}_i(g,f)\right],
\end{equation}
where $\mathcal{R}$ is the scalar curvature, $\Lambda=-\frac{(d-1)(d-2)}{2l^2}$ is a negative cosmological constant and $f$ is a fixed symmetric tensor. $c_i$ is constant and $\mathcal{U}$ are symmetric polynomials of eigenvalues of the $d\times d$ matrix $\mathcal{K}_\nu^\mu=\sqrt{g^{\mu\alpha}f_{\alpha \nu}}$,
\begin{equation} \label{eq:sym}
  \begin{aligned}
    \mathcal{U}_1          & =[\mathcal{K}],                                                                                                                                                                                                                                            \\
    \mathcal{U}_2          & =[\mathcal{K}]^2-[\mathcal{K}^2],                                                                                                                                                                                                                          \\
    \mathcal{U}_3          & =[\mathcal{K}]^3-3[\mathcal{K}][\mathcal{K}^2]+2[\mathcal{K}^3]
    ,                                                                                                                                                                                      \\
        \mathcal{U}_4 & = [ \mathcal { K } ] ^ { 4 } - 6 \left[ \mathcal { K } ^ { 2 } \right] [ \mathcal { K } ] ^ { 2 } + 8 \left[ \mathcal { K } ^ { 3 } \right] [ \mathcal { K } ] + 3 \left[ \mathcal { K } ^ { 2 } \right] ^ { 2 } - 6 \left[ \mathcal { K } ^ { 4 } \right].
  \end{aligned}
\end{equation}
Here, $\mathcal{F}=F_{\mu\nu}F^{\mu\nu}$, where $F_{\mu\nu}$ is the electromagnetic field tensor with $F_{\mu\nu}= \nabla_\mu A_\nu - \nabla_\nu A_\mu$. The parameter $\beta$ is the Born-Infeld parameter; the Born-Infeld field reduces to the linear Maxwell field when $\beta\to \infty$ and vanishes when $\beta\to 0$.

The equation of motion (EOM) of this system is given by,
\begin{equation}
  \label{eq:eom}
  \begin{aligned}
    R_{\mu \nu}-\frac{1}{2}R g_{\mu\nu}+\Lambda g_{\mu \nu}-\frac{1}{2}g_{\mu \nu}L(\mathcal{F})-               & \frac{2F_{\mu \lambda}F^\lambda_\nu}{\sqrt{1+\frac{\mathcal{F}}{2\beta^2}}}+m^2\mathcal{X}_{\mu \nu}=0 ,\\
    \partial_\mu \left ( \frac{\sqrt{-g}F^{\mu \nu}}{\sqrt{1+\frac{\mathcal{F}}{2\beta^2}}}\right ) & =0,
  \end{aligned}
\end{equation}
where $\mathcal{X}_{\mu \nu}$ is the massive term with
\begin{equation}
  \begin{aligned}
    \mathcal{X}_{\mu \nu}= & \frac{c_1}{2}(\mathcal{U}_1g_{\mu \nu}-\mathcal{K}_{\mu \nu})-\frac{c_2}{2}(\mathcal{U}_2g_{\mu \nu}-2\mathcal{U}_1\mathcal{K}_{\mu \nu}+2\mathcal{K}^2_{\mu \nu})-\frac{c_3}{2}(\mathcal{U}_3g_{\mu \nu}-3\mathcal{U}_2\mathcal{K}_{\mu \nu}+      \\
                         & 6\mathcal{U}_1\mathcal{K}^2_{\mu \nu}-6\mathcal{K}_{\mu \nu}^3)-\frac{c_4}{2}(\mathcal{U}_4g_{\mu \nu}-4\mathcal{U}_3\mathcal{K}_{\mu \nu}+12\mathcal{U}_2\mathcal{K}_{\mu \nu}^2-24\mathcal{U}_1\mathcal{K}^3_{\mu \nu}+24\mathcal{K}_{\mu \nu}^4).
  \end{aligned}
\end{equation}
This massive term introduces a mass for the graviton and breaks the diffeomorphism invariance of the bulk gravity theory, which is dual to momentum dissipation in the boundary field theory. We solve the EOM with this ansatz,
\begin{equation} \label{eq:ansatz}
  ds^2=-f(r)dt^2+f^{-1}(r)dr^2+r^2h_{ij}dx_idx_j,\quad i,j=1,2,3,\dots,n,
\end{equation}
where the metric $h_{ij}$ describes a Euclidean space with constant curvature $(d-2)(d-3)k$, and $k$ can be negative (hyperbolic), zero (flat), or positive (elliptic). We consider the reference metric with constant $c_0$
\begin{equation}
  f_{\mu \nu}=diag(0,0,c_0^2h_{ij}).
\end{equation}
The gauge field is $A_\mu=h(r)dt$ and
\begin{equation}
  h(r)=-\sqrt{\frac{d_2}{d_3}}\frac{q}{r^{d_3}}\mathcal{H}.
\end{equation}
In this paper, we set $d_i=d-i$, and $\mathcal{H}$ denotes the hypergeometric function
\begin{equation} \label{eq:hypergeometric_function}
  \mathcal{H}=\sideset{_2}{_1}{\mathop{\mathcal{F}}}\left(\left[\frac{1}{2},\frac{d_3}{2d_2}\right],\left[\frac{3d_{7/3}}{2d_2}\right],-\Gamma\right),
\end{equation}
where $\Gamma=\frac{d_2d_3q^2}{\beta^2r^{2d_2}}$ and $q$ is charge constant. In our ansatz \eqref{eq:ansatz}, $\mathcal{U}_m$ reads,
\begin{equation}
  \begin{aligned}
    \mathcal{U}_1 =\frac{d_2c_0}{r}, \qquad
    \mathcal{U}_2 =\frac{d_2d_3c_0^2}{r^2},  \qquad
    \mathcal{U}_3 =\frac{d_2d_3d_4c_0^3}{r^3},   \qquad
    \mathcal{U}_4 =\frac{d_2d_3d_4d_5c_0^4}{r^4}.  \qquad
  \end{aligned}
\end{equation}
These $\mathcal{U}_m$ expressions enter the black hole metric through the massive term $\mathcal{X}_{\mu\nu}$ and will be used in the subsequent analysis of the metric and entanglement measures. The metric function $f(r)$ is \cite{Hendi:2015hoa}
\begin{equation}
  \begin{aligned}
   f(r)= & k-\frac{m_0}{r^{d_3}}+(\frac{4\beta^2-2\Lambda}{d_1d_2})r^2-\frac{4\beta^2r^2}{d_1d_2}\sqrt{1+\Gamma}+\frac{4d_2q^2\mathcal{H}}{d_1r^{2d_3}} \\
          & +m^2\{\frac{c_0c_1}{d_2}r+c_0^2c_2+\frac{d_3c_0^3c_3}{r}+\frac{d_3d_4c_0^4c_4}{r^2}\}.
  \end{aligned}
\end{equation}
In the metric function, $m_0$ is the mass of the black hole.

\begin{figure}
  \centering
  \includegraphics[width=0.45\textwidth]{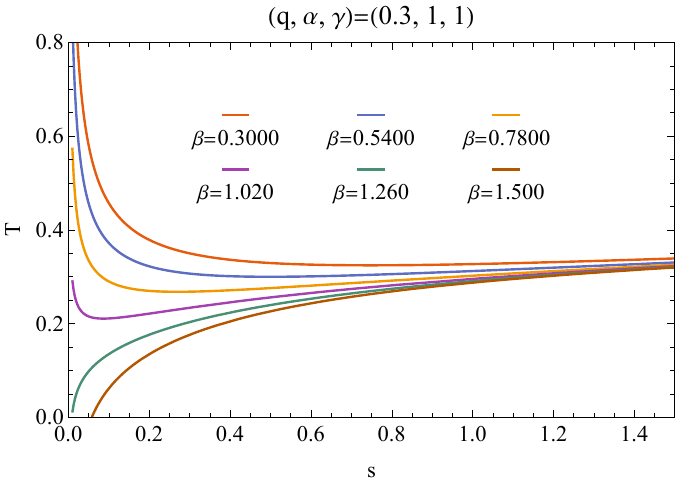}\quad
  \includegraphics[width=0.45\textwidth]{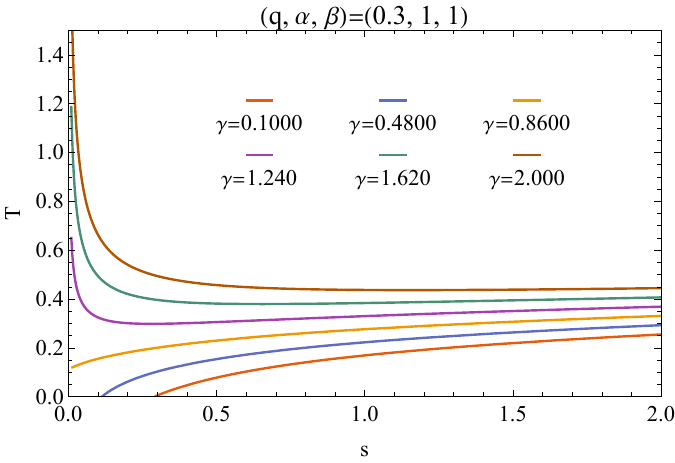}
  \caption{The temperature $T$ versus entropy density $s$ of EBI massive gravity theory with different parameters.}
  \label{fig:pd1}
\end{figure}

In this paper, we only consider 4-dimensional flat space, which means $d=4$ and $k=0$. It is easy to find that $\mathcal{U}_3=\mathcal{U}_4=0$. Furthermore, we set massive term $m^2c_0c_1=\alpha$, $m^2c_0^2c_2=\gamma$, and $c_0=1$ for convenience to calculate. Upon simplification of these functions, we obtain the following expressions:
\begin{equation} \label{eq:black_hole_solution}
  \begin{aligned}
    f(r)=&\gamma +\frac{8 q^2 \mathcal{H}}{3 r^2}-\frac{m_0}{r}-\frac{2}{3} \beta ^2 r^2 \left(\sqrt{\frac{2 q^2}{\beta ^2 r^4}+1}-1\right)+r^2+\frac{\alpha  r}{2}
                                             \\
    \mathcal{H}=                                                  & \sideset{_2}{_1}{\mathop{\mathcal{F}}}\left(\frac{1}{2},\frac{1}{4};\frac{5}{4};-\frac{2q^2}{r^4\beta^2}\right).
  \end{aligned}
\end{equation}
When the radial coordinate equals the horizon radius ($r=r_h$), we have $f(r_h)=0$. So, we can deduce the mass $m_0$ and obtain the Hawking temperature $T$ is
\begin{equation}
  T=\frac{f'(r_h)}{4\pi}=\frac{\gamma +r_h^2 \left(3-2 \beta ^2 \left(\sqrt{\frac{2 q^2}{\beta ^2 r_h^4}+1}-1\right)\right)+\alpha  r_h}{4 \pi  r_h}.
  \end{equation}
Because we only focus on the flat case, the total entropy of the planar black hole is IR-divergent due to infinite transverse area. To address this, we work with the effective entropy density $s = \pi r_h^2$, which is finite. Then we can rewrite the Hawking temperature function:
\begin{equation}
  \label{eq:ts}
  T(s)=\frac{\pi  \gamma +s \left(3-2 \beta ^2 \left(\sqrt{\frac{2 \pi ^2 q^2}{\beta ^2 s^2}+1}-1\right)\right)+\sqrt{\pi } \alpha  \sqrt{s}}{4 \pi ^{3/2} \sqrt{s}}.
\end{equation}
We show the phase diagram of the EBI massive gravity theory in figure \ref{fig:pd1}. From \eqref{eq:ts}, we find massive term $\gamma$ and BI term $\beta$ can determine the behavior of temperature when $s\to 0$.

\subsection{Holographic quantum information}
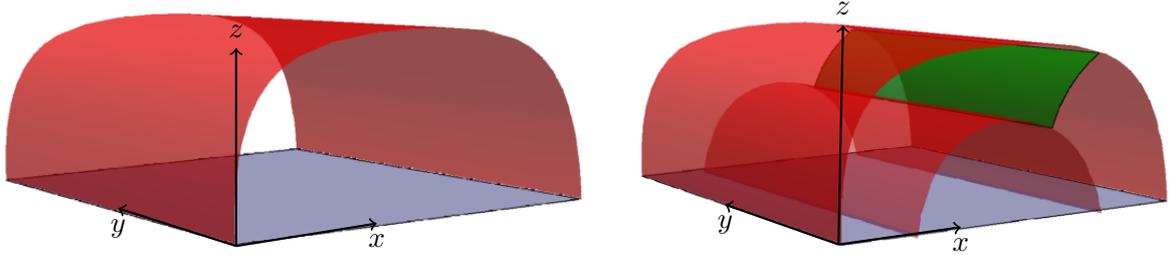
\begin{figure}
  \begin{tikzpicture}[scale=1]
    \node [above right] at (0,0) {\includegraphics[width=8.1cm]{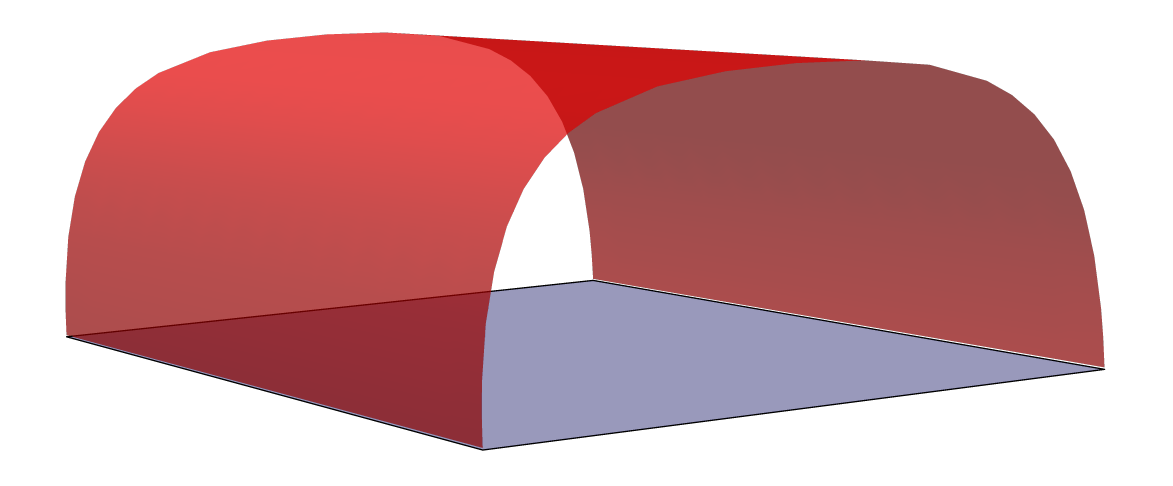}};
    \draw [right,->,thick] (3.43, 0.35) -- (5.25, 0.6) node[below] {$x$};
    \draw [right,->,thick] (3.43, 0.35) -- (1.85, 0.8) node[below] {$y$};
    \draw [right,->,thick] (3.43, 0.35) -- (3.4, 2.825) node[above] {$z$};
  \end{tikzpicture}
  \begin{tikzpicture}[scale=1]
    \node [above right] at (0,0) {\includegraphics[width=7.5cm]{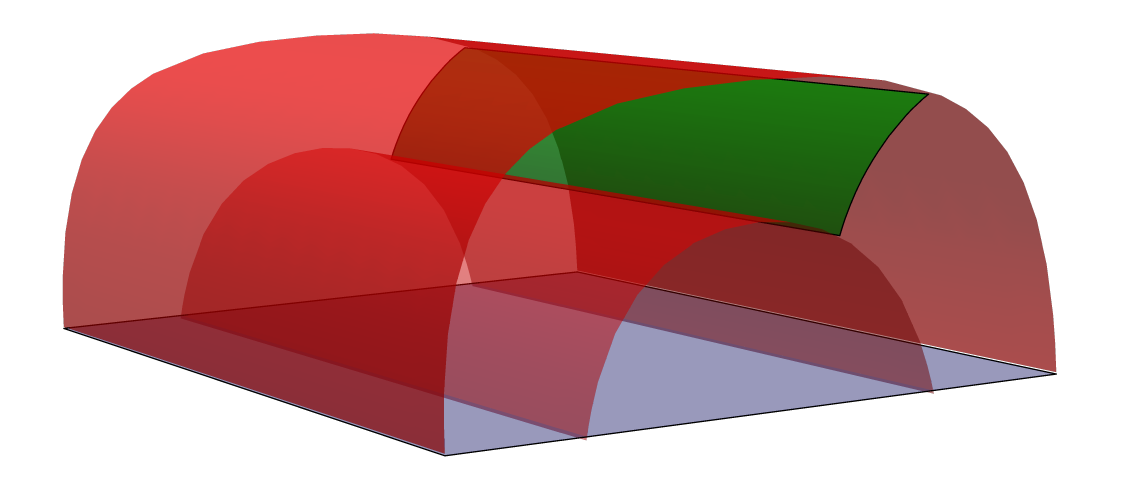}};
    \draw [right,->,thick] (3.08, 0.348) -- (4.55, 0.55) node[below] {$x$};
    \draw [right,->,thick] (3.08, 0.348) -- (1.45, 0.9) node[below] {$y$};
    \draw [right,->,thick] (3.08, 0.348) -- (3.0, 3.125) node[above] {$z$};
  \end{tikzpicture}

  \caption{Left panel: The red surface represents the HEE of the blue subregion with width $w$. Right panel: The red surfaces represent the minimal surfaces, and the green surface represents the EWCS.}
  \label{fig:syhee}
\end{figure}

Quantum information provides a characterization of entanglement. Recently, numerous investigations have studied the holographic duality of quantum information measures. Entanglement entropy (EE) is one of the most widely used entanglement measures, characterizing the entanglement between a subsystem and its complement for pure states. EE is defined in terms of the reduced density matrix $\rho_A$ \cite{Eisert:2008ur},
\begin{equation}
  S_A(|\psi\rangle)=-\text{Tr}[\rho_A\text{log}(\rho_A)],\qquad \rho_A=\text{Tr}_B(|\psi\rangle\langle\psi|).
\end{equation}
The holographic dual of EE corresponds to the minimal surface in the bulk gravitational system, known as HEE \cite{Ryu:2006bv}. In the left panel of figure~\ref{fig:syhee}, we illustrate the schematic of HEE. In this work, we consider a strip configuration extending along the $y$-direction, where the red surface represents the HEE of the dual system. Additionally, the entanglement entropy for a boundary subregion $A$ is computed via the Ryu-Takayanagi formula,
\begin{equation}\label{eq:rt}
   S(A) = \frac{\text{Area}(\gamma_A)}{4G_N},
\end{equation}
 where $\gamma_A$ is the minimal surface anchored to the boundary of $A$ that satisfies the homology condition. In static black hole spacetimes, extremal surfaces cannot probe beyond the event horizon~\cite{Hubeny:2012ry}. Consequently, for the strip subregions considered in this work, the minimal surface $\gamma_A$ remains outside the black hole horizon and does not intersect it, thus trivially fulfilling the homology condition. The subregion configuration of HEE is a strip geometry along the $x$-direction with width $w$, and the corresponding minimal surface is obtained by extremizing the area functional as described in \eqref{eq:rt}. Note that HEE typically diverges due to the asymptotic AdS behavior, and we have regularized HEE by subtracting the divergent term. However, EE is susceptible to classical correlations in mixed-state systems.  Consequently, HEE, as the dual of EE, is often contaminated by thermal entropy when measuring mixed-state entanglement \cite{Ling:2015dma, Ling:2016wyr}. This limitation motivates the search for novel measures capable of capturing the entanglement properties of mixed-state systems.

\begin{figure}
\centering
\includegraphics[width=0.55\textwidth]{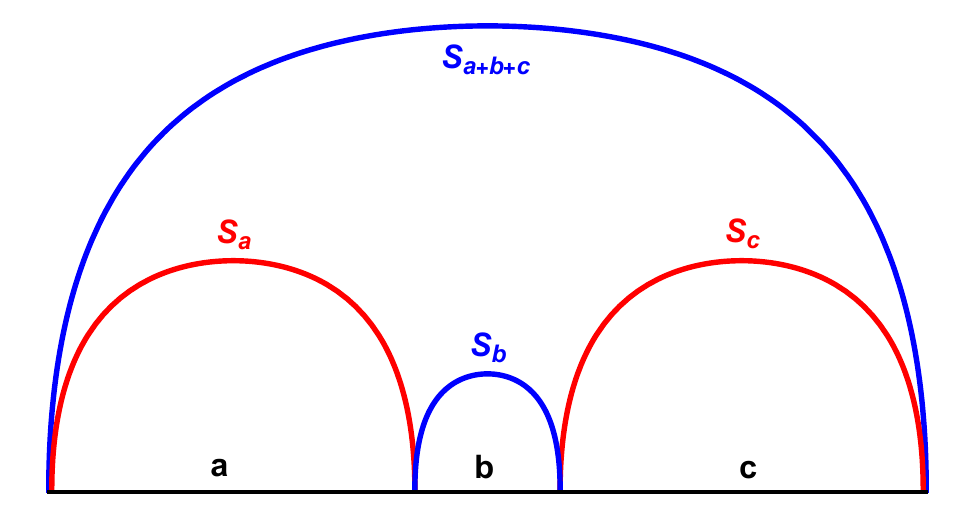}
\caption{The illustration of HMI, the subsystems $a$ and $c$ are separated by the region $b$. The red and blue surfaces represent the HEE of different subsystems.}
\label{fig:misy}
\end{figure}

To better investigate mixed-state entanglement, several mixed-state entanglement measures have been proposed. Among them, the holographic dual of mutual information, referred to as HMI, is one of the most widely used. HMI characterizes the total correlation in a system, encompassing both quantum entanglement and classical correlations. For mixed-state systems, HMI is therefore a more informative correlation measure than HEE, although it does not isolate entanglement alone. For instance, a product state $\rho = \rho_A \otimes \rho_B$ has nonzero von Neumann entropy for each subsystem, yet its HMI vanishes, consistent with the absence of any correlation between the two parties. In our holographic setup, HMI quantifies the total correlation between two spatially separated subsystems $a$ and $c$, with an intermediate region $b$ placed between them. It can be expressed as \cite{chuang:2002, Hayden:2011ag}
\begin{equation}
I(a:c)=S(a)+S(c)-S(a\cup c),
\end{equation}
where $S(x)$ represents the EE of subsystem $x$, and $S(a\cup c)$ is evaluated via the RT prescription, which automatically selects the minimal surface among the connected and disconnected candidate surfaces. Consequently, within the holographic duality framework, HMI is intrinsically linked with the HEE. For HMI, the subsystems $a$ and $c$ are separated by $b$, as depicted in figure \ref{fig:misy}. The red surfaces illustrate the EE for subsystems $a$ and $c$, while the blue surfaces represent the separated region $b$ and the entire region $a+b+c$. However, HMI is also affected by thermal entropy because it is constructed from HEE, so it may not provide a clear characterization of the mixed-state system.

\begin{figure}
  \centering
  \includegraphics[width=0.75\textwidth]{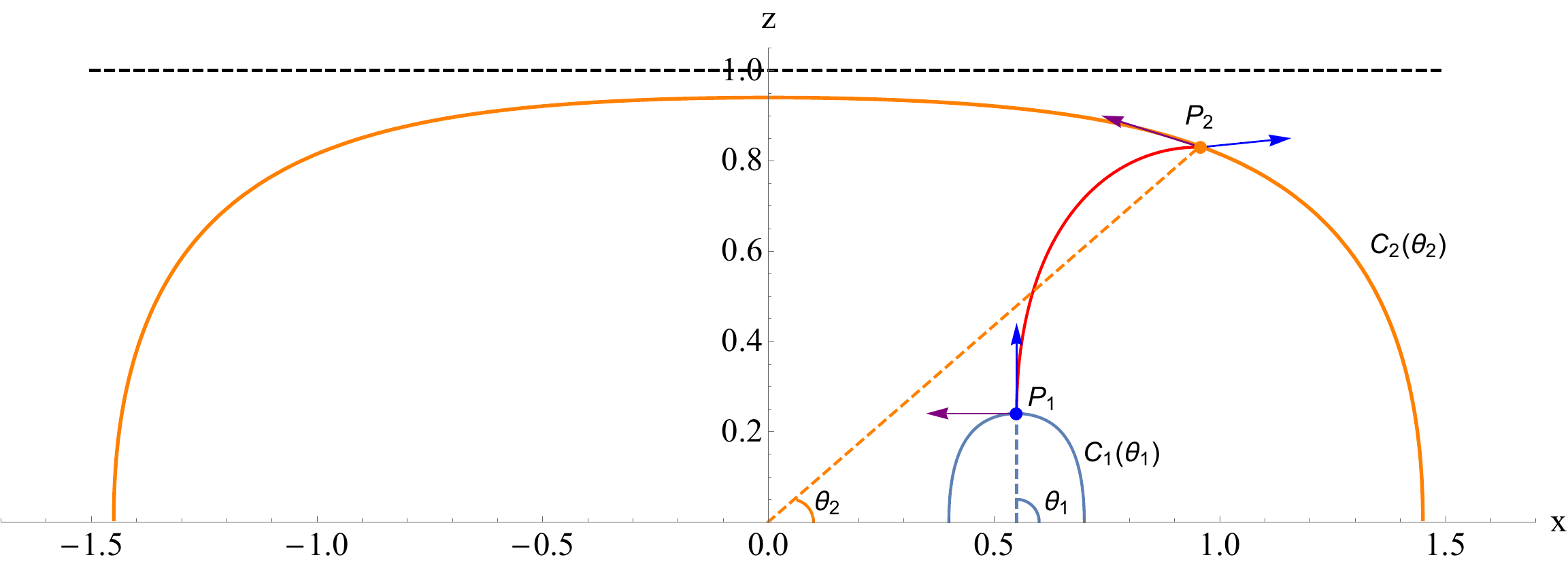}
  \caption{Schematic illustration of the numerical procedure for solving the asymmetric EWCS.}
  \label{fig:algoewcs}
\end{figure}

Recently, EWCS has been proposed as a holographic probe of mixed-state entanglement \cite{Takayanagi:2017knl}. We emphasize that the boundary quantities dual to EWCS are themselves subject to classical correlations at finite temperature. Nevertheless, EWCS accesses the entanglement through the entanglement wedge, a geometric structure distinct from the RT minimal surface that determines HEE and HMI, so it may carry entanglement information that HEE and HMI do not capture. The definition of EWCS is given by,
\begin{equation}
  E_w(\rho_{AB})=\underset{\Sigma_{AB}}{\text{min}}\left( \frac{\text{Area}(\Sigma_{AB})}{4G_N}  \right),
\end{equation}
where $\Sigma_{AB}$ represents the cross-section in the entanglement wedge that ends on the boundary of $A$ and $B$. Many studies show that EWCS can be thought of as the dual of reflected entropy, logarithmic negativity and odd entropy \cite{Kudler-Flam:2018qjo, Dutta:2019gen, Jokela:2019ebz, BabaeiVelni:2019pkw, Vasli:2022kfu, Camargo:2022mme, Tamaoka:2018ned, Nguyen:2017yqw}. It should be noted that reflected entropy, although widely used in holographic studies of mixed-state entanglement, is not a measure of physical correlations. It fails to be monotonically decreasing under partial trace, and none of the R\'enyi reflected entropies with $0<\alpha<2$ qualifies as a correlation measure, even for classical probability distributions \cite{Hayden:2023yij}. Since the precise boundary dual of EWCS remains an open question, we do not attribute the properties of any single candidate dual to EWCS itself. In this work, EWCS is used as a bulk geometric quantity, namely the minimal cross-section of the entanglement wedge, which probes mixed-state entanglement through the geometric changes of the wedge. Our conclusions follow from its geometric temperature dependence rather than from the correlation-measure properties of a particular boundary dual. In this paper, we consider the EWCS for a system with subsystems $a$ and $c$ separated by $b$. We show the illustration of EWCS in the right panel of figure~\ref{fig:syhee}. The entanglement wedge is the region between two red minimal surfaces, and the green surface represents EWCS. In addition to this, EWCS only occurs when the entanglement wedge is exist, which means HMI does not vanish. The computation of the EWCS is highly sensitive to the choice of bipartite configuration. While symmetric setups are relatively tractable, asymmetric configurations are more physically relevant yet pose significant computational difficulties. The primary challenge lies in the need to search over a two-dimensional parameter space to identify the minimal entanglement-wedge cross-section, a procedure that is numerically demanding. Furthermore, conventional coordinate systems become singular near the AdS boundary, compromising numerical stability and precision. In light of these difficulties, we employ the efficient and robust numerical scheme developed in \cite{Liu:2019qje} for computing the asymmetric EWCS. A schematic outline of the algorithm is provided in figure~\ref{fig:algoewcs}.

The background metric is expressed as
  \begin{equation}
    ds^2=g_{tt}dt^2+g_{zz}dz^2+g_{xx}dx^2+g_{yy}dy^2.
  \end{equation}
  Let $C_1(\theta_1)$ and $C_2(\theta_2)$ denote the two minimal surfaces that form the connected configuration, each anchored to the cross-section slice at points $P_1$ and $P_2$, respectively. The area of
  the cross-section then reads
  \begin{equation}
    A=\int_{C_{P_1}C_{P_2}}\sqrt{g_{xx}g_{yy}x'(z)^2+g_{zz}g_{yy}}dz.
  \end{equation}
  The equation of motion governing the locally minimal cross-section is derived as
  \begin{equation}
    x'(z)^3\left (\frac{g_{xx}g_{yy}'}{2g_{yy}g_{zz}}+\frac{g_{xx}'}{2g_{zz}}\right )+x'(z)\left (\frac{g_{xx}'}{g_{xx}}+\frac{g_{yy}'}{2g_{yy}}-\frac{g_{zz}'}{2g_{zz}}\right )+x''(z)=0.
  \end{equation}
  Since the minimal cross-section must be locally orthogonal to the boundary of the entanglement wedge, we impose
  \begin{equation}\label{eq:orth}
    Q_1(\theta_1,\theta_2) \equiv\left.\frac{\langle \frac{\partial}{\partial z},\frac{\partial}{\partial \theta_1}\rangle}{\sqrt{\langle \frac{\partial}{\partial z},\frac{\partial}{\partial z}\rangle
  \langle\frac{\partial}{\partial \theta_1},\frac{\partial}{\partial \theta_1}\rangle}} \right |_{p_1}=0,\quad
    Q_2(\theta_1,\theta_2) \equiv\left.\frac{\langle \frac{\partial}{\partial z},\frac{\partial}{\partial \theta_1}\rangle}{\sqrt{\langle \frac{\partial}{\partial z},\frac{\partial}{\partial z}\rangle
  \langle\frac{\partial}{\partial \theta_2},\frac{\partial}{\partial \theta_2}\rangle}} \right |_{p_2} =0,
  \end{equation}
  where $\langle\cdot,\cdot\rangle$ represents the inner product induced by the bulk metric $g_{\mu\nu}$. Once Eq.~\eqref{eq:orth} is satisfied, the minimal cross-section is anchored on the minimal surface at
  the parameters $(\theta_1,\theta_2)$. To locate the endpoints $(P_1,P_2)$, we adopt a Newton–Raphson iteration scheme combined with a pseudospectral method. Using this algorithm, we proceed to evaluate the
  asymmetric EWCS in the EBI massive gravity theory, providing a quantitative probe of mixed-state entanglement in this system.

Employing the efficient algorithm from \cite{Liu:2019qje} described above, we are able to investigate more properties of mixed-state entanglement. Using these mixed-state entanglement measures, we conduct a comprehensive investigation into the intricate relationship between the effective MIT and the Hawking-Page phase transition. This approach allows us to quantify the entanglement properties that emerge during these distinct phase transitions, potentially revealing novel insights into the underlying quantum nature of these phenomena.

\section{The relationship between the transport properties and mixed-state entanglement}
\label{sec:mit}
The variation of transport properties is one of the most important topics in condensed matter, often associated with phase transitions in the system \cite{RevModPhys.79.677, VENUGOPAL201229, C0NR00403K}. Within the holographic duality framework, the investigation of transport properties assumes importance. Recent studies have extensively explored various aspects of transport properties, including electronic and thermoelectric conductivity \cite{Donos:2014cya, Donos:2014uba}. Furthermore, several holographic models have been employed to investigate the transport phenomena \cite{Baggioli:2016oju, Baggioli:2014roa, Vegh:2013sk, Zeng:2014uoa, Baggioli:2021xuv, Gouteraux:2016wxj}. These investigations have significantly advanced our understanding of condensed matter systems, particularly in the context of strongly correlated electron systems. In the EBI massive gravity model, we have examined the DC conductivity of the system and identified the occurrence of an effective MIT with temperature. More importantly, near the crossover temperature of effective MIT, we studied the mixed-state entanglement measures. This discovery can help us deeply investigate the relationship between the mixed-state entanglement and the nature of phase transitions.

\subsection{The DC conductivity of the system}
In this paper, we use the ansatz in \eqref{eq:ansatz} to calculate the DC conductivity \cite{Donos:2014uba}. We can solve the EOM of the gauge field and get
\begin{equation}
 q=\frac{-r^2 h'(r)}{\sqrt{1-\frac{ h'(r)^2}{ \beta ^2}}}.
\end{equation}
Therefore, we can consider switching on an electric field $E$ on the gauge field $A_x$. The perturbation of the ansatz reads,
\begin{equation}
A_x=-E t+\delta h(r),\quad
g_{tx}=r^2\delta g_{tx}(r),\quad
g_{rx}=r^2\delta g_{rx}(r).
\end{equation}
With this perturbation ansatz, we can solve \eqref{eq:eom} and algebraically solve $\delta g_{rx}$ as
\begin{equation}
  \delta g_{tx}(r)=-\frac{2 \text{E} h'(r)}{\sqrt{1-\frac{h'(r)^2}{\beta ^2}} \left(-\gamma +\frac{1}{2} r^2 f''(r)+r f'(r)+2 \beta ^2 r^2 \sqrt{1-\frac{h'(r)^2}{\beta ^2}}-2 \beta ^2 r^2-\alpha  r\right)}.
\end{equation}
\begin{figure}
  \centering
  \includegraphics[width=0.45\textwidth]{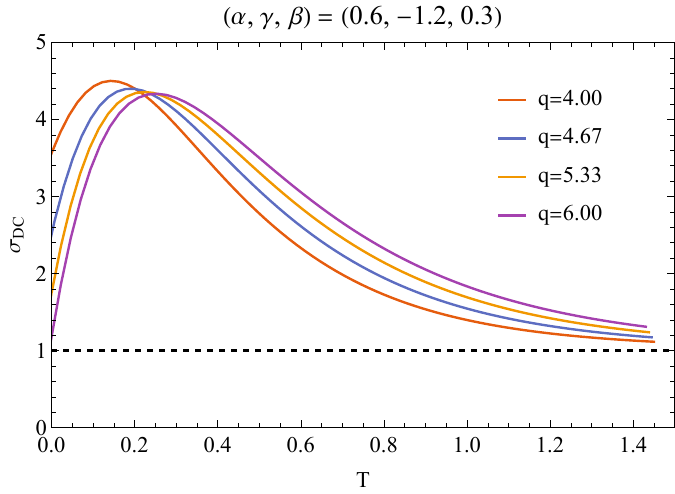}\quad
  \includegraphics[width=0.45\textwidth]{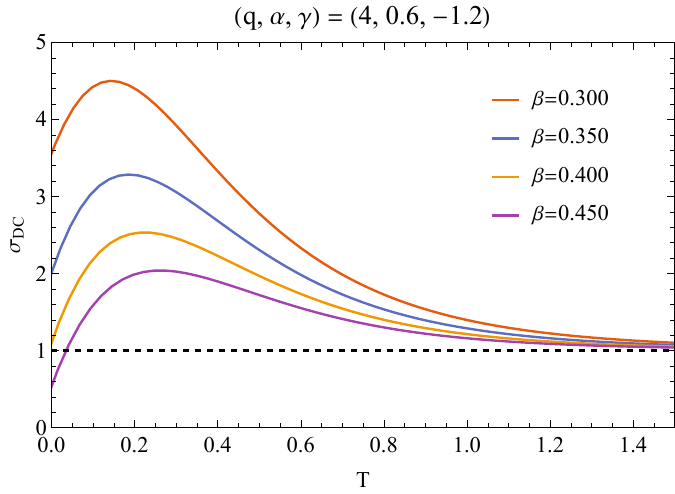}
  \caption{DC conductivity $\sigma_{\mathrm{DC}}$ versus temperature $T$. Left panel: Parameters $(\alpha,\gamma,\beta)=(0.6,-1.2,0.3)$ with different $q$. Right panel: Parameters $(q,\alpha,\gamma)=(4,0.6,-1.2)$ with different BI parameter $\beta$.}
  \label{fig:dc1}
  \end{figure}
Additionally, we can obtain the current $j$ with this perturbation ansatz
\begin{equation}
j=\frac{-r^2 \delta g_{tx}(r) h'(r)-f(r) \delta h'(r)}{\sqrt{1-\frac{h'(r)^2}{\beta ^2}}}.
\end{equation}
The DC conductivity can be obtained by $\sigma_{\mathrm{DC}}=j/E$ and evaluated at the horizon of the black hole ($r=r_h$). Therefore, the DC conductivity $\sigma_{\mathrm{DC}}$ of this system reads,
\begin{equation}
 \sigma_{\mathrm{DC}}=\frac{1}{\sqrt{1-\frac{h'(r_h)^2}{\beta ^2}}} \left( 1+\frac{2 r_h^2 h'(r_h)^2}{ \left(-\gamma +\frac{1}{2} r_h^2 f''(r_h)+r_h f'(r_h)+2 \beta ^2 r_h^2 \sqrt{1-\frac{h'(r_h)^2}{\beta ^2}}-2 \beta ^2 r_h^2-\alpha  r_h\right)}\right).
\end{equation}

From the above equation, we can study the DC conductivity of the EBI massive gravity model. The varying behavior of $\sigma_{\mathrm{DC}}$ can signify the different phases of the model. It is straightforward to derive the function $\sigma_{\mathrm{DC}}(T)$, where the insulating phase is characterized by $\sigma_{\mathrm{DC}}'(T)>0$ and the metallic phase is determined by $\sigma_{\mathrm{DC}}'(T)<0$ \cite{Ling:2015exa, Ling:2015dma, Donos:2014cya}. Under certain specific parameters of this model, the effective MIT can occur as the temperature increases. In figure \ref{fig:dc1}, we illustrate the effective MIT with varying parameters $q$ and $\beta$. As the temperature rises, the insulating phase initially appears at lower temperatures and subsequently transitions to the metallic phase at higher temperatures.

Near the crossover temperature of effective MIT, the different phases not only have different transport properties but also could have different entanglement behavior. In the next section, we will introduce the behavior of different entanglement measures when the phase transition occurs.

\subsection{The mixed-state entanglement measures of effective MIT}\label{sec:effmitewcs}

\begin{figure}
\centering
\includegraphics[width=0.41\textwidth]{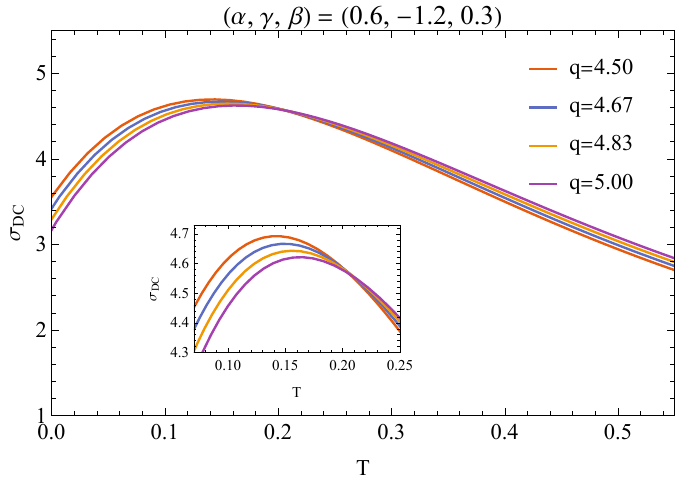}\quad
\includegraphics[width=0.45\textwidth]{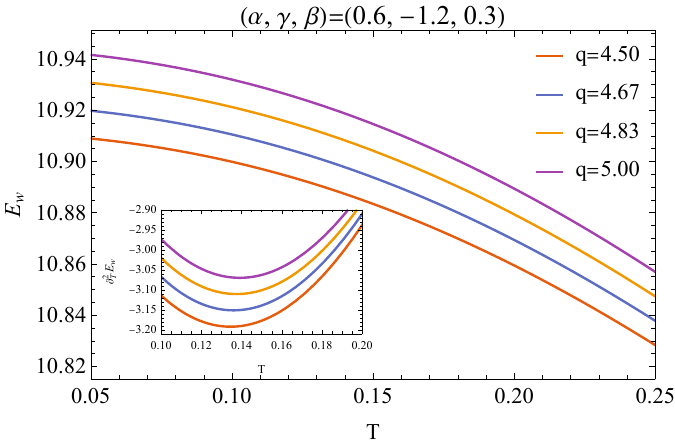}
\caption{Behavior of parameters $\alpha=0.6, \gamma=-1.2, \beta=0.3$ with different $q$ when effective MIT occurs. Left panel: DC conductivity $\sigma_{\mathrm{DC}}$ versus temperature $T$. The inset shows the phase transition region of effective MIT. Right panel: EWCS $E_w$ versus temperature $T$ with configuration $(a, b, c)=(0.5, 0.1, 0.5)$. The inset shows the second-order partial derivative of $E_w$ with respect to temperature.}
\label{fig:ewvt1}
\end{figure}

As discussed in Sec.~\ref{sec:introduction}, EWCS is a mixed-state entanglement probe in the holographic duality framework. In the Born-Infeld (BI) theory, a coupling between the entanglement structure and the transport behavior has been discussed \cite{Liu:2023rhd}. Consequently, the exploration of this relationship is necessary. In figure \ref{fig:ewvt1}, we delve into the behavior of EWCS during the effective MIT at a specific temperature. Our findings indicate that EWCS exhibits a monotonically decreasing trend with the rise in temperature. However, the second-order partial derivative of EWCS with respect to temperature presents a non-monotonic pattern. Intriguingly, the region exhibiting this non-monotonic behavior closely aligns with the crossover temperature of effective MIT. From a boundary perspective, this non-monotonic behavior can be understood naturally if we treat EWCS as a holographic probe of mixed-state entanglement, without committing to a specific boundary dual. The candidate duals of EWCS, including entanglement of purification, reflected entropy, odd entropy, and logarithmic negativity, capture different but related aspects of the entanglement structure~\cite{Kudler-Flam:2018qjo, Dutta:2019gen, Jokela:2019ebz, BabaeiVelni:2019pkw, Vasli:2022kfu, Camargo:2022mme, Tamaoka:2018ned, Nguyen:2017yqw}. As the boundary theory transitions from the insulating side ($\sigma_{\mathrm{DC}}'(T) > 0$) to the metallic side ($\sigma_{\mathrm{DC}}'(T) < 0$), this entanglement structure is reorganized, and the temperature dependence of EWCS tracks this reorganization. From the geometric perspective, this reorganization has a concrete origin. The entanglement wedge whose cross-section defines EWCS is bounded by the RT surfaces of the two subsystems, and the cross-section is anchored on these surfaces. For large subsystem configurations, these RT surfaces extend close to the black hole horizon and are dominated by the thermal entropy, so that EWCS inevitably inherits the infrared geometry of the thermal state. The temperature dependence of $E_w$ is thus a geometric property of the entanglement wedge itself, rather than a statement about any specific boundary dual. The monotonic decrease of $E_w$ with temperature reflects this geometric restructuring along the crossover. The peak of the second-order derivative $\partial^2 E_w/\partial T^2$ is located close to the crossover temperature, with a small parameter-dependent offset quantified in Appendix~\ref{appendixA}. Therefore, the higher-order terms of EWCS provide a geometric diagnostic of the effective MIT, in addition to the conductivity-based probe.

\begin{figure}
  \centering
  \includegraphics[width=0.41\textwidth]{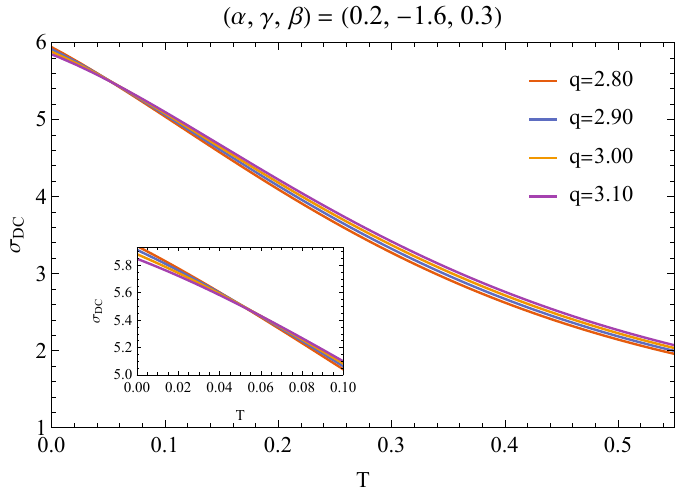}\quad
  \includegraphics[width=0.45\textwidth]{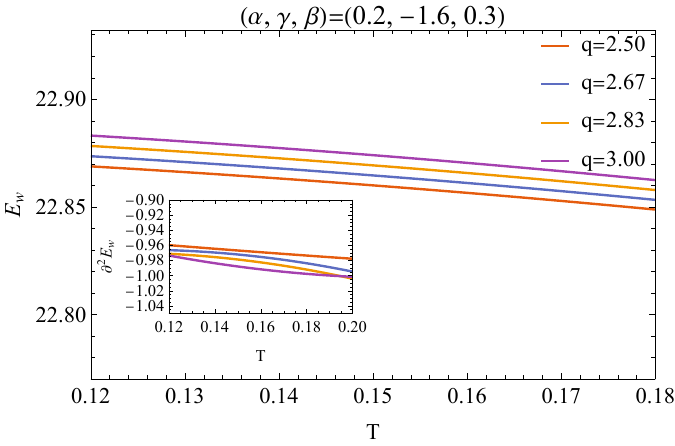}
  \caption{The behavior of parameters $\alpha=0.2, \gamma=-1.6, \beta=0.3$ with different $q$ when effective MIT does not occur. Left panel: DC conductivity $\sigma_{\mathrm{DC}}$ versus temperature $T$. The inset shows the first-order derivative of $\sigma_{\mathrm{DC}}$ with respect to temperature. Right panel: EWCS $E_w$ versus temperature $T$ with configuration $(a, b, c)=(0.5, 0.05, 0.45)$. The inset shows the second-order partial derivative of $E_w$ with respect to temperature.}
  \label{fig:ewvt2}
  \end{figure}

\begin{figure}
\centering
\includegraphics[width=0.45\textwidth ]{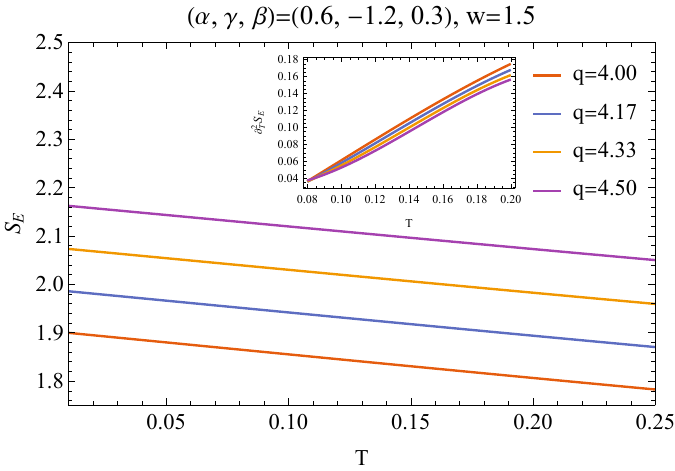}\quad
\includegraphics[width=0.45\textwidth]{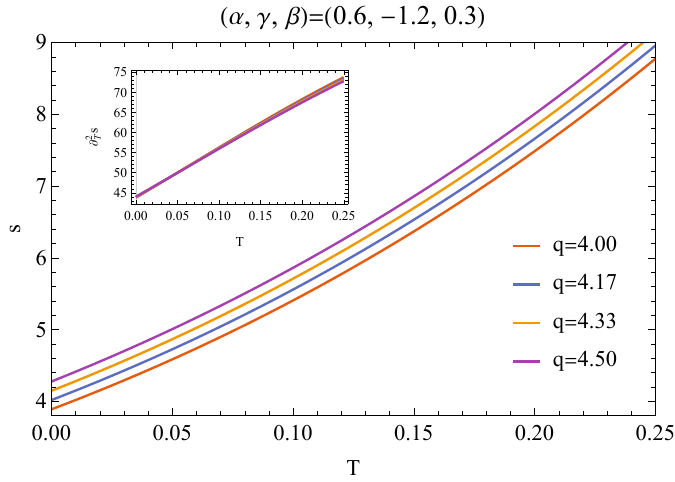}
\caption{Behavior of parameters $\alpha=0.6$, $\gamma=-1.2$, and $\beta=0.3$ when effective MIT occurs. Left panel: HEE $S_E$ versus temperature $T$. The inset shows the second-order partial derivative of $S_E$ with respect to temperature. Right panel: Entropy density $s$ versus temperature $T$. The inset shows the second-order partial derivative of $s$ with respect to temperature.}
\label{fig:heevt1}
\end{figure}

In addition to the aforementioned analysis, we have also examined the relationship between EWCS and DC conductivity in the absence of an effective MIT phase transition. As depicted in figure \ref{fig:ewvt2}, the $\sigma_{\mathrm{DC}}'(T)$ is less than $0$, indicating it is always in the metallic phase. However, the EWCS and the second-order derivative of EWCS with respect to temperature exhibit monotonic behavior when the effective MIT phase transition is absent. This behavior suggests that the non-monotonic behavior is exclusive to instances where the effective MIT phase transition occurs. Furthermore, we have compared the behavior of the mixed-state entanglement probe with that of other holographic entanglement measures. For instance, HEE is a widely used measure of entanglement, but it is notably susceptible to the influence of thermal entropy \cite{Yang:2023wuw, Liu:2023rhd, Ling:2015dma}. In figure \ref{fig:heevt1}, we present the behavior of HEE during an effective MIT phase transition. Both HEE and its second-order derivative with respect to temperature display monotonic behavior as the temperature increases. Moreover, the large width of HEE is determined by the thermal entropy, and the behavior of $S_E$ is similar to the entropy density $s$. Both the entropy density $s$ and HEE $S_E$ exhibit monotonic behavior, and their higher-order terms are consistently monotonic.

\begin{figure}
\centering
\includegraphics[width=0.47\textwidth]{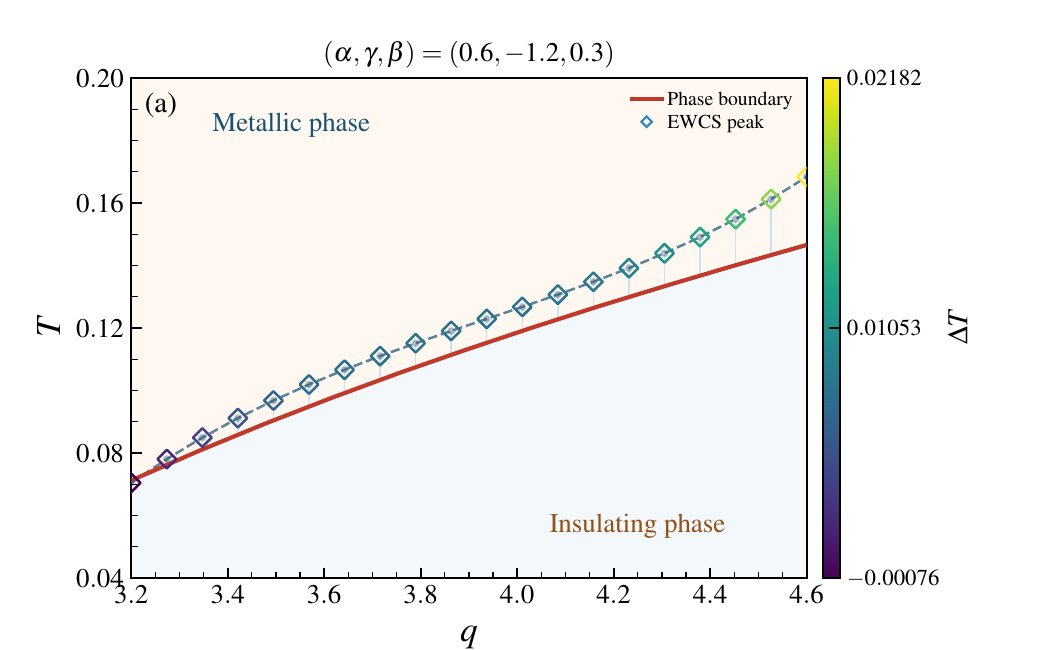}\quad
\includegraphics[width=0.47\textwidth]{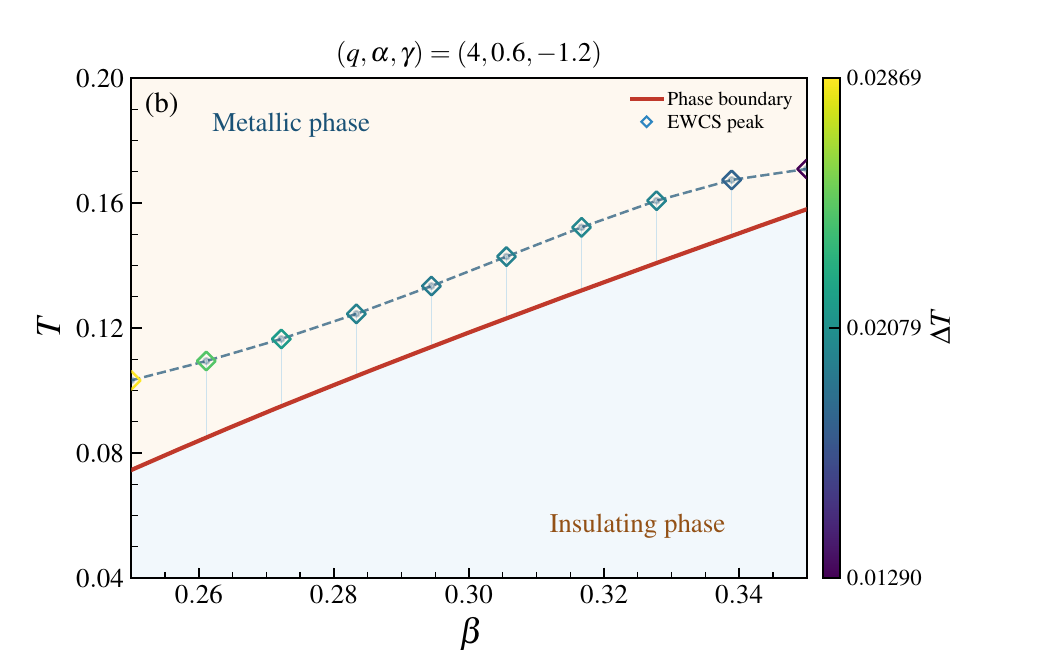}\quad
\includegraphics[width=0.47\textwidth]{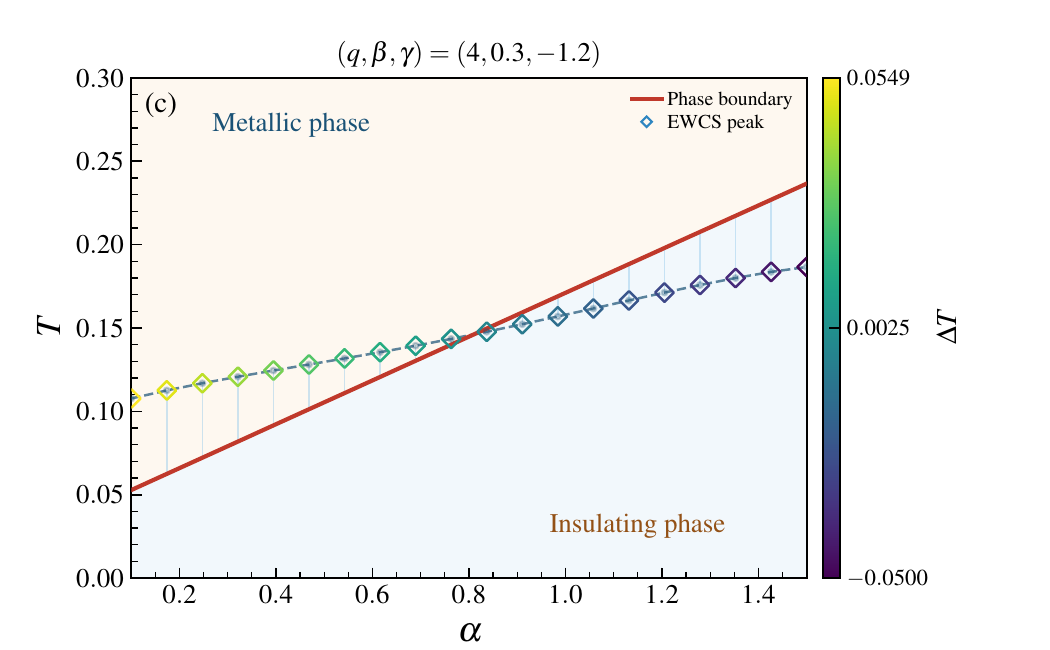}\quad
\includegraphics[width=0.47\textwidth]{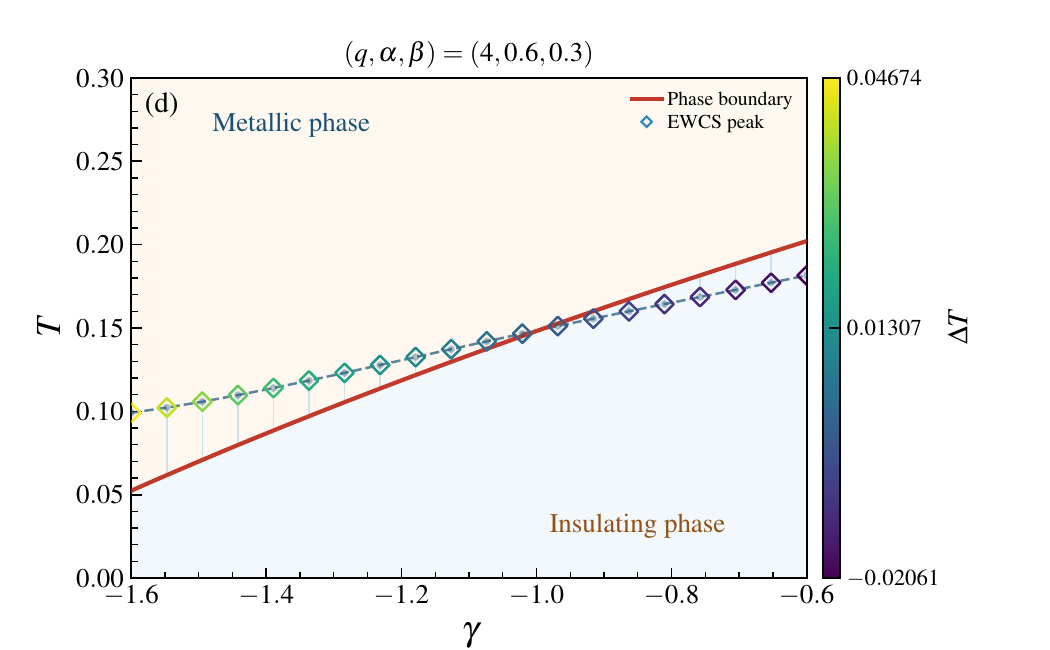}\quad
\caption{The relationship between the peak of the EWCS with configuration $(a, b, c)=(0.5, 0.1, 0.5)$ and the phase diagram of effective MIT. The red triangle represents the peak of the higher-order term of EWCS. Panel (a): The phase diagram of charge $q$ versus temperature $T$. Panel (b): The phase diagram of BI parameter $\beta$ versus temperature $T$. Panel (c): The phase diagram of massive term $\alpha$ versus temperature $T$. Panel (d): The phase diagram of massive term $\gamma$ versus temperature $T$.}
\label{fig:phase}
\end{figure}

To gain a more comprehensive understanding of the relationship between EWCS and effective MIT, we have also examined the phase diagram of effective MIT alongside the peak of the higher-order terms of EWCS, as shown in figure \ref{fig:phase}. It is important to note that not all regions of the parameter space in EBI massive gravity theory are physically admissible, as some parameter spaces include imaginary parts in the bulk. Within the valid region of the parameters, including charge $q$, Born-Infeld parameter $\beta$, massive parameters $\alpha$ and $\gamma$, the peak of the higher-order terms of EWCS is consistently located near the crossover temperature. A systematic statistical analysis of the correlation between the EWCS peak trajectory and the crossover temperature, including the quantitative deviation $\Delta T$ and its scaling across all four parameter directions, is presented in Appendix~\ref{appendixA}. This observation also suggests a potential correlation between EWCS and effective MIT, which appears to be concealed within the higher-order terms of the mixed-state entanglement probe. This correlation could provide valuable insights into the phase transitions and entanglement measures, thereby contributing to our understanding of holographic dualities.

  In the EBI massive gravity model, we have investigated the behavior of entanglement measures during the effective MIT. The higher-order derivatives of the mixed-state entanglement probe can detect phase transitions at finite temperatures, whereas HEE shows no such feature. Similar behavior has been observed in lattice studies. Quantum discord, in contrast to entanglement and thermodynamic quantities, detects a quantum critical point even at finite temperature~\cite{Werlang:2010xds}, in analogy with EWCS detecting the thermal crossover. The area-law coefficient of the R\'enyi entanglement negativity is singular across a finite-temperature transition~\cite{Wu:2019qvm}. While HEE is dominated by thermal entropy at finite temperatures, the mixed-state entanglement probe accesses the entanglement structure through a geometric structure complementary to that of the RT minimal surface. These results indicate that this probe provides information distinct from that of HEE. The enhanced sensitivity of this mixed-state entanglement probe to the MIT crossover suggests its potential as a complementary diagnostic tool for entanglement analysis. Since phase transitions at finite temperatures are ubiquitous in condensed matter systems, this probe could have broad applications in such situations. Additionally, the Hawking-Page phase transition also occurs in this model, which we explore in the following section.

\section{The relationship between entanglement measurements and Hawking-Page phase transition}
\label{sec:hawkingpage}

\begin{figure}
  \centering
  \includegraphics[width=0.45\textwidth]{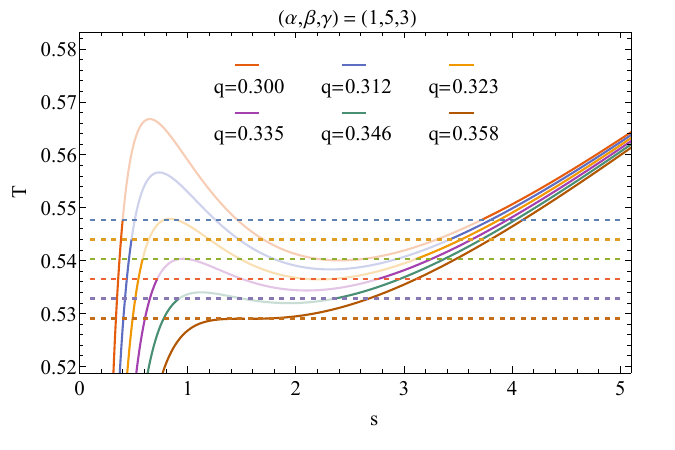}\quad
  \includegraphics[width=0.44\textwidth]{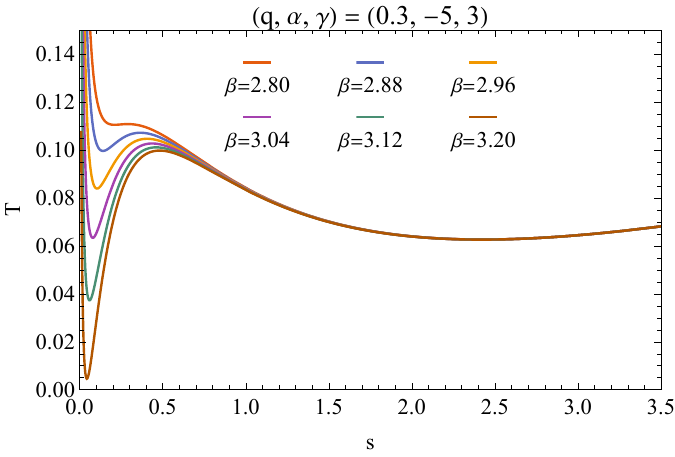}
  \caption{Phase diagram of temperature $T$ versus entropy density $s$ during the Hawking-Page phase transition. Left panel: First-order and second-order phase transitions with different values of charge $q$. Stable and metastable states are shown. Dashed lines indicate critical temperatures for different parameters. The brown line ($q=0.358$) represents the second-order phase transition. Right panel: A special first-order phase transition with different BI parameters $\beta$.}
  \label{fig:pd2}
\end{figure}

\begin{figure}
  \includegraphics[width=0.45\textwidth]{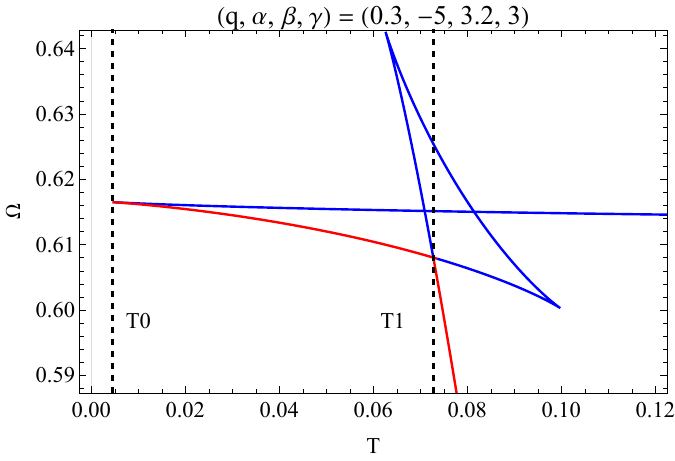}\quad
  \includegraphics[width=0.45\textwidth]{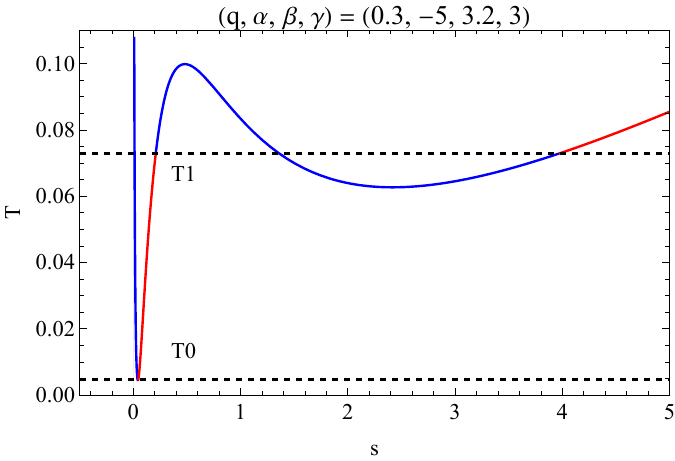}
  \caption{A special case of Hawking-Page phase transition in EBI massive gravity theory. Left panel: Free energy $\Omega$ versus temperature $T$. Stable and metastable states are shown. The first-order phase transition occurs at critical temperature $T_1=0.0728$.
  Right panel: Temperature $T$ versus entropy density $s$. Stable and metastable states are shown. The system has a minimum temperature $T_0=0.0455$.}
  \label{fig:special_free}
\end{figure}

For specific parameter values, the system can undergo a Hawking-Page phase transition. The entropy density $s$ is associated with the horizon $r_h$ of the black hole. In figure \ref{fig:pd2}, we present the phase diagram featuring both first-order and second-order phase transitions. The left figure illustrates both first-order and second-order phase transitions, where the black hole can possess multiple horizons at the same temperature. As the temperature increases, the horizon abruptly transitions from one to another at the critical temperature. For a typical first-order phase transition, there are three different horizons $r_h$ at the same temperature. However, the right figure \ref{fig:pd2} depicts a distinctive type of first-order phase transition, where there are four different horizons at the same temperature. To more effectively investigate the phase transition in this special case, we define the free energy as $\Omega= M- T s$, where $M$ represents the mass of the black hole, $T$ is the temperature, and $s$ is the entropy density. Consequently, the free energy can be expressed as follows
\begin{equation}
\Omega = \frac{8 q^2 \, \sideset{_2}{_1}{\mathop{\mathcal{F}}}\left(\frac{1}{4},\frac{1}{2};\frac{5}{4};-\frac{2 q^2}{r_h^4 \beta ^2}\right)}{3 r_h}+\frac{1}{12} r_h \left(9 \gamma +r_h^2 \left(3-2 \beta ^2 \left(\sqrt{\frac{2 q^2}{\beta ^2 r_h^4}+1}-1\right)\right)+3 \alpha  r_h\right).
\end{equation}
In figure~\ref{fig:special_free}, we show the free energy and the phase transition for this special case. Only the red lines represent stable states, while the blue lines represent metastable states. The first-order phase transition occurs at the critical temperature $T_1=0.0728$. In this special case, the system has a minimum temperature $T_0=0.0455$. This phase possesses a finite minimum temperature $T_0$ and can have three different metastable horizons at certain temperatures, which differs from the normal first-order phase transition shown in the left panel of figure~\ref{fig:pd2}.

\subsection{The Holographic entanglement entropy and mutual information}

\begin{figure}
  \centering
  \includegraphics[width=0.45\textwidth]{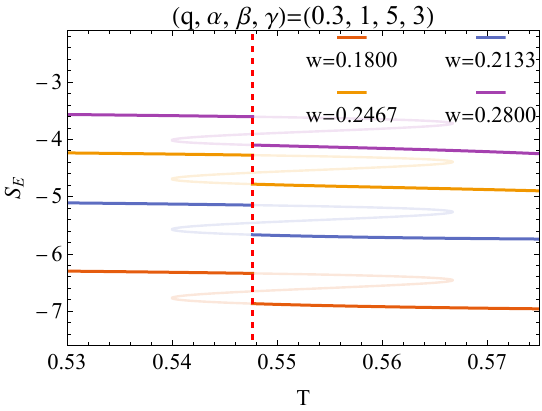}\quad
  \includegraphics[width=0.45\textwidth]{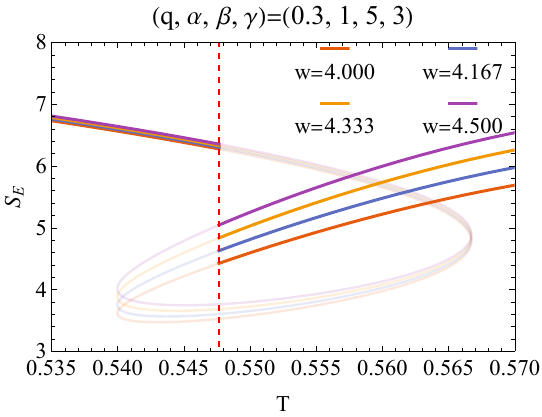}
  \caption{HEE $S_E$ versus temperature $T$ for several different widths during a first-order phase transition. Stable and metastable states are shown. The red dashed line indicates critical temperature $T_c=0.5476$. HEE exhibits a jump with increasing temperature near the critical point. Left panel: HEE behavior for small widths. Right panel: HEE behavior for large widths.}
  \label{fig:hee}
\end{figure}

When we set the parameters $(q,\alpha,\beta,\gamma)=(0.3,1,5,3)$, the system undergoes a first-order phase transition. As depicted in figure \ref{fig:hee}, the behavior of HEE is related to the configuration. We observe that HEE decreases with increasing temperature in a small-width configuration, and it abruptly jumps when the temperature reaches the critical point. We note that this decreasing behavior with temperature for small widths appears to contradict the conventional field-theoretic expectation that entanglement entropy grows with temperature. The resolution lies in the different regions probed by the RT surface. For small widths, the RT surface probes only the near-boundary geometry and remains far from the horizon region, so the thermal contribution is suppressed. In the present model, the Hawking-Page transition is realized as a small/large black-hole transition in the temperature-entropy relation, and the entanglement measures inherit its branch structure. Below $T_c$, the system stays on the stable small-black-hole branch, along which the bulk solution evolves monotonically with temperature; the narrow-strip HEE is computed on this branch solution and follows this evolution, decreasing monotonically as $T$ rises toward $T_c$. This monotonic decrease does not contradict the field-theoretic expectation that entanglement entropy grows with temperature: that expectation is a conformal statement, derived for a unique thermal state family, where the small-width entropy grows with temperature, with a positive leading thermal correction~\cite{Calabrese:2004ch}. The present model is not conformal, since the massive-gravity parameters $\alpha$ and $\gamma$ deform the near-boundary geometry, and the same small-width decrease has been found in massive gravity~\cite{Liu:2021rks}, where HEE is known to mirror the thermal phase structure~\cite{Zeng:2015tfj}. At $T_c$, the first-order transition abruptly changes the horizon structure, producing the observed jump. For large widths, by contrast, the thermal entropy contribution takes over after the transition, and HEE rises together with the entropy density $s(T)$ above $T_c$. In the right panel of figure \ref{fig:hee}, we illustrate that HEE exhibits increasingly complex behavior as the width of the configuration increases. Initially, HEE decreases until it reaches the critical temperature, beyond which it experiences a rapid increase with rising temperature. Notably, the critical behavior of HEE closely mirrors that of the entropy density in configurations with large width $w$. The entropy density $s$ demonstrates a similar positive correlation with temperature. This phenomenon can be attributed to the dominance of thermal entropy in HEE calculations for large configurations \cite{Ling:2015dma}.

\begin{figure}
  \centering
  \includegraphics[width=0.45\textwidth]{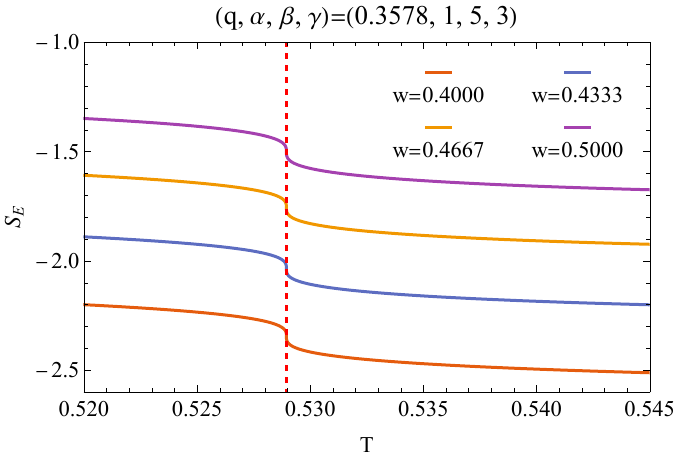}\quad
  \includegraphics[width=0.45\textwidth]{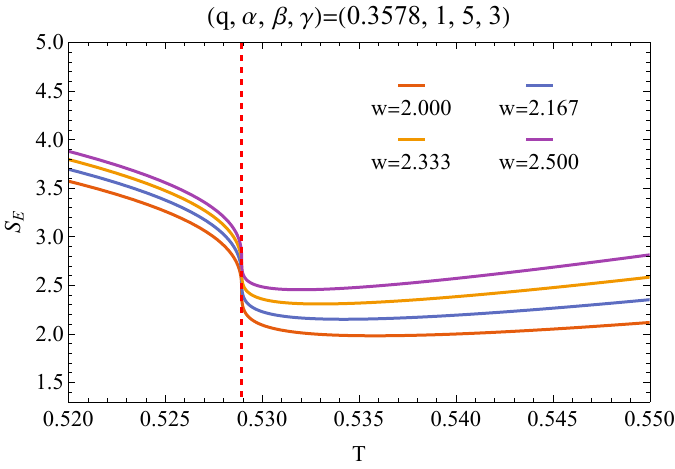}
  \caption{HEE $S_E$ versus temperature $T$ with parameters $(q,\alpha,\beta,\gamma)=(0.3578,1,5,3)$. The red dashed line indicates the critical temperature $T_c=0.5289$ of the second-order phase transition. Left panel: HEE behavior for small widths. Right panel: HEE behavior for large widths.}
  \label{fig:heesecond}
\end{figure}

In figure \ref{fig:heesecond}, we depict the behavior of HEE in the second-order phase transition. In the case of a small width, HEE decreases as the temperature increases. However, for a large width, HEE increases with the rise in temperature upon crossing the critical point. The behavior of HEE in a large width is similar to that of a first-order phase transition. This suggests that similar to the first-order phase transition, HEE in a large-width configuration is easily influenced by the thermal entropy. This consistency across different phase transitions underscores the significant role of thermal entropy in shaping the behavior of HEE, particularly in large-width configurations. The small-width decrease of HEE in the second-order transition shares the same physical origin as the first-order case: HEE follows the evolution of the stable branch, and for small widths the RT surface stays in the near-boundary region far from the horizon. Since the second-order transition involves a continuous change of the horizon rather than a jump, HEE now decreases smoothly and exhibits the power-law singular behavior discussed in Sec.~\ref{sec:sca}, in contrast to the discontinuous jump of the first-order case.

\begin{figure}
  \centering
  \includegraphics[width=0.45\textwidth]{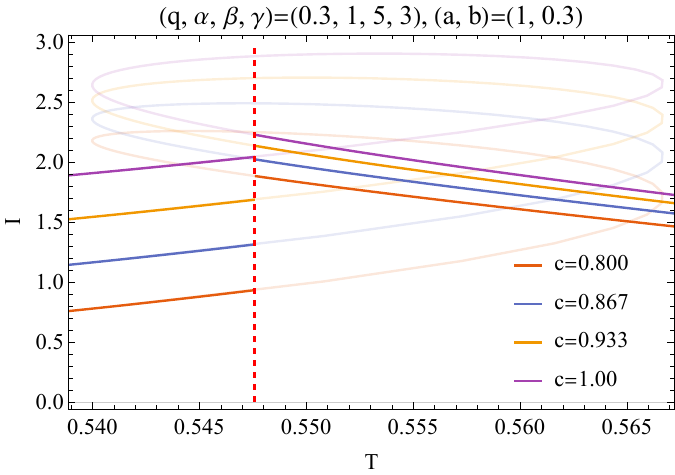}\quad
  \includegraphics[width=0.45\textwidth]{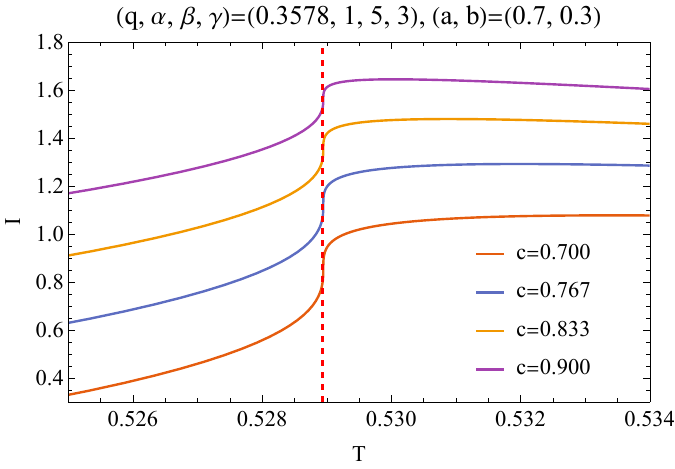}
  \caption{Holographic mutual information $I$ versus temperature $T$. The red dashed lines indicate the critical temperature. Left panel: HMI versus temperature for different configurations during a first-order phase transition with critical temperature $T_c=0.5476$. Right panel: HMI versus temperature for different configurations during a second-order phase transition with critical temperature $T_c=0.5289$.}
  \label{fig:mi}
\end{figure}

Contrary to the HEE, HMI can also be utilized to quantify the total correlation in a system \cite{Hayden:2011ag}. As depicted in figure \ref{fig:mi}, the behavior of HMI as a function of temperature is illustrated. Our findings indicate that HMI exhibits a behavior opposite to that of HEE for large widths. Initially, HMI increases until it reaches a critical point, after which it decreases with increasing temperature. This pattern suggests that HMI behaves oppositely to HEE at large widths, where HEE is dictated by thermal entropy. Consequently, it can be inferred that HMI can also be influenced by thermal entropy \cite{Huang:2019zph, Liu:2021rks}. The temperature profile of HMI requires a physical explanation, since mutual information is conventionally expected to decrease monotonically with temperature. Note that the regularized HEE is negative, since the divergent part is subtracted, so a decrease of HEE with temperature means that its magnitude grows. Since $I(a:c)=S(a)+S(c)-S(a\cup c)$, the temperature derivative of HMI is $dI/dT=|S(a\cup c)|'-|S(a)|'-|S(c)|'$ in terms of the magnitudes of the subsystems. Below $T_c$, all HEE decrease as $T$ approaches the critical point, and the union $a\cup c$ decreases faster than each of the two subsystems, because its entanglement is more sensitive to the change of the stable horizon branch. In our results, the growth rate of the union exceeds the sum of the two subsystem rates, so that $dI/dT>0$ and HMI increases as $T$ approaches $T_c$ from below. At $T_c$, the first-order transition abruptly increases the horizon, producing the feature of HMI at the critical point, and above $T_c$ the thermal entropy contribution dominates so that HMI reverts to its conventional decreasing profile. For the second-order transition the same competition operates through the continuous transition of the horizon.

We have examined the behavior of HEE and HMI within the EBI massive gravity theory. These entanglement measures can potentially serve as diagnostic tools for the Hawking-Page phase transition. More specifically, both HEE and HMI exhibit discontinuous and singular behavior near the critical temperature, indicative of first-order and second-order phase transitions. Furthermore, we find that the behavior of HEE is related to the width. At large widths, HEE is dictated by thermal entropy, exhibiting behavior similar to entropy density. However, while the definition of HMI is inherently related to HEE, it consistently exhibits an inverse relationship with HEE at large widths. This suggests that HMI, being constructed from HEE, does not by itself provide a complete characterization of mixed-state entanglement. As discussed in Sec.~\ref{sec:effmitewcs}, the boundary duals of EWCS carry the same classical-correlation caveat as HEE and HMI. Nevertheless, EWCS accesses the entanglement structure through the entanglement wedge, a geometric structure distinct from the RT minimal surface. Although this structure is anchored on the RT surfaces and cannot completely exclude the influence of the infrared geometry, it is not determined by the thermal entropy alone, unlike large-width HEE. It is therefore valuable to also examine this complementary probe alongside HEE and HMI.

\subsection{Entanglement wedge cross-section}

\begin{figure}
  \centering
  \includegraphics[width=0.45\textwidth]{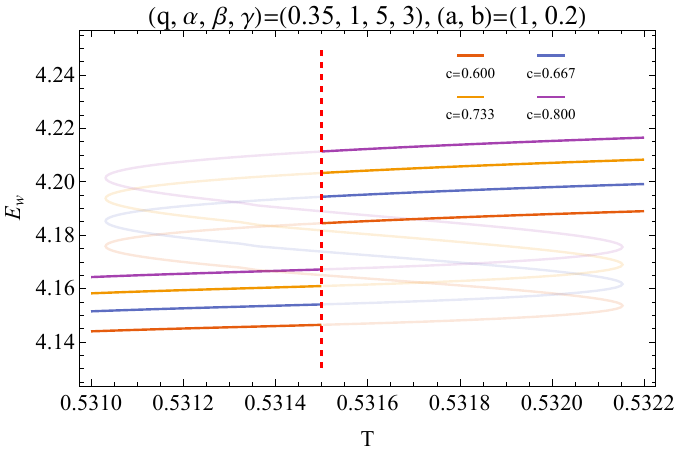}
  \includegraphics[width=0.45\textwidth]{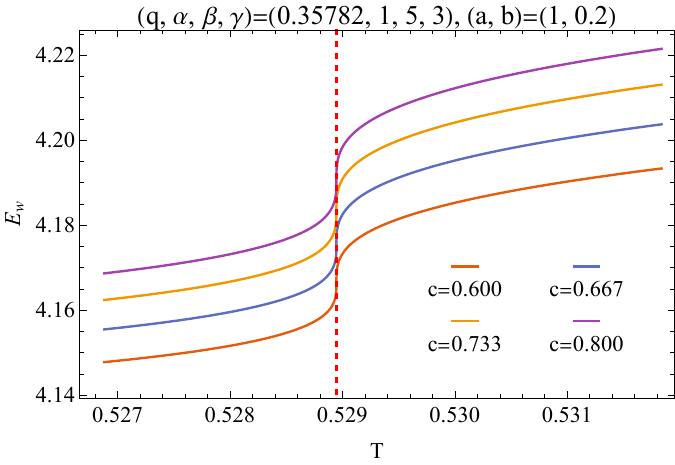}
  \caption{EWCS $E_w$ versus temperature $T$ for several different configurations. Left panel: EWCS behavior during a first-order phase transition. Stable and metastable states are shown. The red dashed line indicates the critical temperature $T_c=0.5315$. Right panel: EWCS behavior during a second-order phase transition. The red dashed line indicates the critical temperature $T_c=0.528948$.}
  \label{fig:eop}
\end{figure}

The relationship between EWCS and temperature $T$ is depicted in figure~\ref{fig:eop}. Our findings reveal that EWCS also exhibits phase transition behavior near the critical point of the Hawking-Page phase transition. As the temperature $T$ increases, EWCS monotonically increases and experiences an abrupt jump near the critical temperature associated with the first-order phase transition. Notably, this behavior of EWCS differs from that of HMI and HEE. Furthermore, for the Hawking-Page phase transition, EWCS is independent of the strip configuration in EBI massive gravity theory.

In the subsequent analysis, we focus on the second-order phase transition as depicted in figure \ref{fig:eop}. Our findings reveal that EWCS increases monotonically with rising temperature and exhibits singularity near the critical temperature. As in the first-order phase transition, the mixed-state entanglement probe behaves differently from HMI and HEE. Moreover, the relationship between EWCS and temperature remains independent of any specific configuration.

To summarize the behavior of EWCS in EBI massive gravity theory, we highlight two key points. First, EWCS can effectively diagnose both first-order and second-order phase transitions. It exhibits an abrupt jump and singularity near the critical temperatures of first-order and second-order phase transitions, respectively. Second, as discussed in Sec.~\ref{sec:introduction}, EWCS is a mixed-state entanglement probe, even though its precise boundary dual remains an open question. Within the Hawking-Page transition in the EBI massive gravity theory, the mixed-state entanglement probe increases monotonically with temperature, in contrast to HEE and HMI. We find that the non-monotonic behavior of HEE and HMI is associated with large subregion configurations, where the influence of thermal entropy cannot be neglected. Consequently, HEE and HMI are dominated by thermal entropy on the high-temperature side of the phase transition. In contrast, EWCS shows a monotonically increasing behavior, which reflects its access to the entanglement structure through a geometric structure different from the RT minimal surface. These distinctive features suggest that EWCS provides diagnostic information beyond that of HEE and HMI, making it a valuable probe of mixed-state entanglement. Furthermore, EWCS holds potential for application in the study of phase transitions across a wide range of theories.

\subsection{The scaling behavior of Hawking page phase transition}
\label{sec:sca}

In the preceding section, we investigated various holographic quantum information measures in the EBI massive gravity theory, specifically focusing on HEE, HMI, and EWCS. These quantum information measures are connected to the geometry of the dual spacetimes. Our analysis reveals that at the second-order critical point of this model, these holographic quantum information measures exhibit scaling behavior. The study of scaling behavior is crucial for our research as it facilitates the classification of different types of phase transitions. Understanding the relationships between various scaling behaviors can provide valuable insights into holographic quantum information.

The function between temperature $T$ and entropy density $s$ is given by \eqref{eq:ts}. In figure \ref{fig:pd2}, we show that in the vicinity of the critical point associated with the second-order phase transition, the derivative of the entropy density exhibits singular behavior,
\begin{equation}
s'(T)\to\infty.
\end{equation}
Therefore, the scaling behavior between entropy density $s$ and temperature $T$ is
\begin{equation}
(s-s_c)\sim(T-T_c)^{\alpha_s},
\end{equation}
where $\alpha_s$ is the critical exponent, $T_c$ is the critical temperature, and $s_c$ is the critical entropy density. In the EBI massive gravity, the analytical calculation of the critical exponent presents significant challenges. This complexity arises primarily from the non-linear terms introduced by the Born-Infeld field. The first derivative of the entropy density with respect to temperature can be expressed as
 \begin{equation}\label{eq:st}
   s'(T)=\frac{8 \pi ^{3/2} \Xi s^{5/2}}{s^2 \left(2 (\Xi-1) \beta ^2+3 \Xi\right)-\pi  \Xi \gamma  s+4 \pi ^2 q^2},
 \end{equation}
 where $\Xi=\sqrt{\frac{2 \pi ^2 q^2}{\beta ^2 s^2}+1}$. However, the intricate nature of this function renders further analytical treatment of its coefficients particularly difficult. Despite the analytical complexity, the phase diagram illustrated in figure \ref{fig:pd2} offers an alternative approach to determining the critical exponent, avoiding the need for intricate analysis. By examining the function $T(s)$ at the second-order phase transition, we observe that both $T'(s)=0$ and $T''(s)=0$ at the critical point. This observation implies that the first two terms in the expansion of $T(s)$ vanish at the critical point, leaving only the third-order term as the dominant contribution. Consequently, in the vicinity of the critical point,  $T(s)$ can be expressed as
\begin{equation}
  T - T_c \propto (s - s_c)^3.
\end{equation}
 This relationship between temperature and entropy density in the critical point leads to the conclusion that the critical exponent for this system is
\begin{equation}
\alpha_s = \frac{1}{3}.
\end{equation}

This exponent originates from the saddle-node bifurcation structure of the solution space. At the critical point, the leading term of the expansion is $T-T_c \propto (s-s_c)^3$. In the EBI massive gravity theory, the BI parameter $\beta$ modifies the expansion coefficients through the nonlinear term $\Xi=\sqrt{\frac{2\pi^2 q^2}{\beta^2 s^2}+1}$. The same exponent $1/3$ is shared by every geometric quantity that depends smoothly on the metric, consistent with the universal scaling observed in pure massive gravity~\cite{Liu:2021rks}.

\begin{figure}
  \centering
  \includegraphics[width=0.45\textwidth]{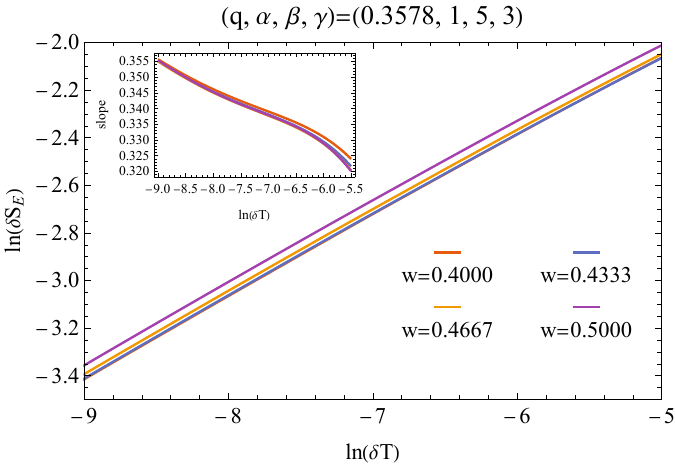}\quad
  \includegraphics[width=0.45\textwidth]{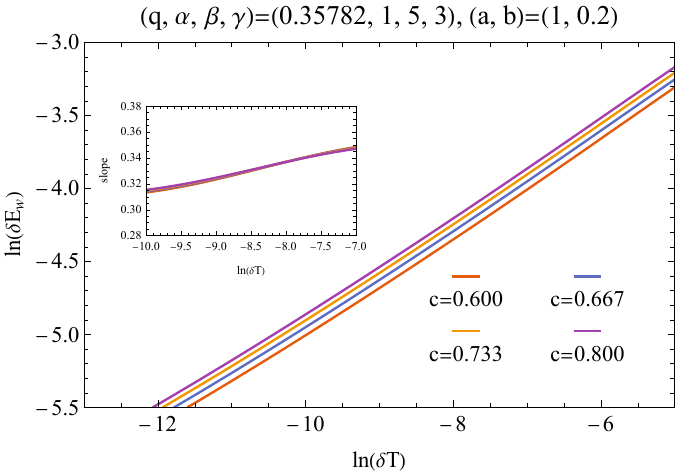}
  \caption{Scaling behavior of different entanglement measures, with slopes converging to $1/3$. Left panel: Log-log plot of HEE deviation $\delta S_E$ versus $\delta T$. Right panel: Log-log plot of EWCS deviation $\delta E_w$ versus $\delta T$.}
  \label{fig:sc}
\end{figure}

Furthermore, it is worth noting that the holographic entanglement measurement also exhibits scaling behavior during the occurrence of second-order phase transitions. To further investigate the scaling behavior, the geometry-related physics quantity $A$ can be expressed as
\begin{equation}
  A=A_c+A' \delta g_{\mu \nu}.
  \label{eq:core}
\end{equation}
From the critical behavior of entropy density, we find that
\begin{equation}
  \delta g_{\mu \nu} \sim (T-T_c)^{\alpha_s}
\end{equation}
To facilitate the calculation of critical exponents, we set
\begin{equation}
  \delta S_E\sim \left(1-\frac{T}{T_c}\right)^{\alpha_{\text{HEE}}},\quad \delta E_w\sim \left(1-\frac{T}{T_c}\right)^{\alpha_{\text{EWCS}}},
\end{equation}
where $\delta S_E \equiv S_E - S_c$ and $\delta E_w \equiv E_w - E_{w_c}$. Figure \ref{fig:sc} illustrates the scaling behavior of both HEE and EWCS in the vicinity of the critical point, which reveals that both $\delta S_E$ and $\delta E_w$ exhibit a power-law relationship with $\delta T$ in the vicinity of the critical point. The log-log plots of these relationships converge to a slope of $1/3$, indicating that the critical exponent of the HEE and EWCS is
\begin{equation}
  \alpha_{\text{HEE}} = \alpha_{\text{EWCS}} = \frac{1}{3}.
\end{equation}
Notably, the same scaling behavior has been observed in massive gravity theory \cite{Liu:2021rks}.

In our investigation of the EBI massive gravity model, we have examined the scaling behavior of the system during the Hawking-Page phase transition. Our analysis demonstrates that when the phase diagram exhibits characteristics of a second-order phase transition, as illustrated in figure~\ref{fig:pd2}, the system consistently manifests a critical exponent of $\alpha_s=1/3$. Moreover, all holographic quantum information measures share this same critical exponent.

\section{Discussion} \label{sec:discuss}

In this paper, we have investigated several holographic entanglement measures, including HEE, HMI, and EWCS, in EBI massive gravity theory. This theory exhibits distinct phase transitions, namely the effective MIT and the Hawking-Page phase transition.

For the finite-temperature effective MIT, we identified EWCS as a mixed-state entanglement probe. The peak of the higher-order derivative of EWCS aligns closely with the crossover temperature of the phase transition. This behavior is unique to EWCS, as the higher-order derivatives of HEE show no such feature. This distinction arises because HEE is dominated by thermal entropy at finite temperature~\cite{Ling:2015dma}. The enhanced sensitivity of EWCS can be understood from both boundary CFT and holographic geometry perspectives.. From the boundary CFT perspective, each candidate boundary dual of EWCS, including entanglement of purification, reflected entropy, odd entropy, and logarithmic negativity, captures a distinct aspect of the entanglement structure~\cite{Kudler-Flam:2018qjo, Dutta:2019gen, Tamaoka:2018ned}. These boundary quantities are themselves subject to classical correlations at finite temperature. Nevertheless, the higher-order derivatives of the mixed-state entanglement probe track the effective MIT crossover. This indicates that the transport crossover in a strongly coupled CFT with momentum dissipation is accompanied by a reorganization of the entanglement structure. This effect is not visible in thermal entropy or in HEE. The peak of $\partial^2 E_w/\partial T^2$ is located close to the crossover temperature, as shown in Appendix~\ref{appendixA}, and therefore serves as a geometric diagnostic of the effective MIT. We emphasize that these findings should be interpreted as properties of the entanglement wedge itself. The precise boundary dual of EWCS remains an open question. In this work, EWCS is used as a bulk geometric quantity that probes mixed-state entanglement through the geometric changes of the entanglement wedge. Reflected entropy, one of the candidate boundary duals, is known not to be a measure of physical correlations~\cite{Hayden:2023yij}. This caveat, however, concerns the candidate dual rather than the cross-section itself, so it does not affect our conclusions. Our conclusions concern the temperature dependence of a bulk geometric quantity and its ability to locate the phase transition, rather than the correlation-measure properties of a particular boundary dual. From the holographic geometry perspective, HEE's RT surface inevitably wraps a portion of the black hole horizon for large subregions, absorbing thermal entropy. EWCS probes the cross-section of the entanglement wedge without directly contacting the horizon~\cite{Hubeny:2012ry}. As explained in Sec.~\ref{sec:effmitewcs}, the entanglement wedge is bounded by the RT surfaces, so EWCS inevitably inherits the infrared geometry. Nevertheless, the cross-section probes the interior of the wedge, a structure complementary to the RT surfaces. EWCS therefore provides entanglement information that HEE and HMI do not capture.

A key finding is the universal critical exponent $\alpha_s = \alpha_{\text{HEE}} = \alpha_{\text{EWCS}} = 1/3$ near the second-order critical point. This exponent arises from the saddle-node bifurcation structure: at a second-order critical point, the cubic expansion $T-T_c \propto (s-s_c)^3$ is a mathematical consequence of the saddle-node condition $T'(s_c)=T''(s_c)=0$, independent of the specific gravitational theory. This universality explains why the same exponent appears in both pure massive gravity~\cite{Liu:2021rks} and our EBI model. The value of the EBI model lies in demonstrating the robustness of this exponent across a broader class of theories that include nonlinear electrodynamics and momentum dissipation simultaneously. We close by summarizing the physics behind the width-dependent and temperature-dependent behavior of HEE and HMI, which seems to deviate from the standard field-theoretic expectation that entanglement increases with temperature. Entanglement measures probe different regions of the geometry, depending on the width. HEE decreases with temperature at small widths, while at large widths it is governed by thermal entropy after the transition and increases with temperature. The crossover is set by the width at which the RT surface begins to approach the horizon. HMI is related to the HEE, so it inherits the competition between entanglement and thermal entropy. It increases as the critical point is approached from below and reverts to a decreasing profile once thermal entropy dominates above $T_c$. These features enhance the diagnostic power of holographic mixed-state probes for finite-temperature phase transitions.

In the EBI model, the combined nonlinearities generate a richer phase structure, including a novel first-order transition with four coexisting horizons and a special transition with finite minimum temperature $T_0$, which is absent in either subsystem alone. The four-horizon structure originates from the BI nonlinearity creating additional extrema in $T(s)$ via the $\sqrt{1+2 q^2/(\beta^2 r^4)}$ term. In such a multi-branch thermodynamic landscape, the EWCS may exhibit hysteresis effects, where the difference $\Delta E_w$ between cooling and heating branches would quantify the entanglement hysteresis of the system.

Our finding that EWCS higher-order derivatives track the MIT crossover point provides a geometric diagnostic distinct from conventional transport-based probes. Recent work connecting EWCS to transport coefficients provides complementary theoretical support~\cite{Li:2024emah}. Related studies of quantum criticality report critical scaling of mixed-state probes with model-dependent exponents. For example, $\alpha\approx 0.7$ for EWCS in a holographic superconductor-insulator transition~\cite{Yang:2026sit}. These results suggest that the critical scaling of mixed-state probes is generic though model-dependent.

Having established that EWCS is an effective probe for finite-temperature transitions, a natural extension is to explore its behavior at zero temperature. Since quantum phase transitions are associated with renormalization group flow between different infrared fixed points, and since EWCS accesses the entanglement structure through the entanglement wedge, EWCS may prove to be a valuable probe for such transitions. We anticipate that EWCS will be a valuable tool for characterizing quantum critical phenomena and are actively pursuing this direction.

\section*{Acknowledgments}

Peng Liu would like to thank Yun-Ha Zha, Yi-Er Liu and Bai Liu for their kind encouragement during this work. Z.Y.\ would like to thank his family and Feng-Ying Deng for their encouragement. This work is supported by the Natural Science Foundation of China under Grant Nos. 12475054, 12035016, 12275275, 12375055 and the Guangdong Basic and Applied Basic Research Foundation No. 2025A1515012063. Z.Y.\ is supported by the Jiangsu Postgraduate Research and Practice Innovation Program under Grant No.\ KYCX25\_3922.

\appendix
\section{Correlation Between EWCS Peak and Crossover Temperature}
\label{appendixA}
\begin{figure}
  \centering
  \includegraphics[width=0.9\textwidth]{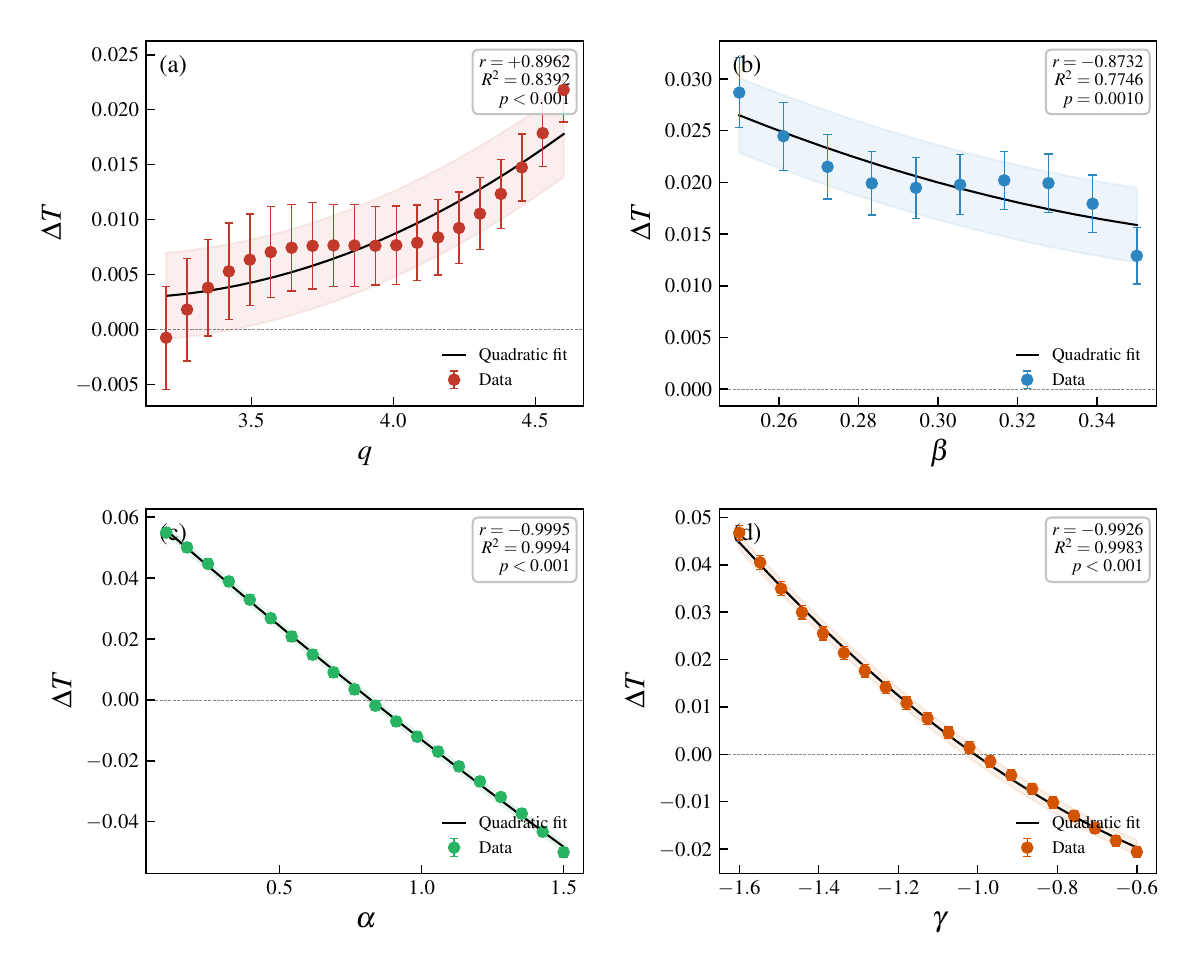}
  \caption{EWCS peak deviation $\Delta T$ as a function of (a) $q$, (b) $\beta$, (c) $\alpha$, and (d) $\gamma$. The data points (circles) are shown with error bars representing the local phase-boundary interpolation uncertainty. The black solid curve is a quadratic fit, and the shaded band denotes the $95\%$ confidence interval. The Pearson correlation coefficient $r$, coefficient of determination $R^2$, and permutation-test $p$-value are displayed in each panel.}
  \label{fig:deviation_analysis}
  \end{figure}

To quantitatively establish the relationship between the EWCS peak position and the crossover temperature, we define the deviation
\begin{equation}
  \Delta T^i \equiv T_{\mathrm{EWCS}}^i - T_{\mathrm{critical}}^i,
\end{equation}
where $i$ represents the different model parameter ($q$, $\beta$, $\alpha$, $\gamma$), $T_{\mathrm{EWCS}}^i$ is the temperature at the EWCS peak, and $T_{\mathrm{critical}}^i$ is the crossover temperature of the effective MIT. In figure~\ref{fig:deviation_analysis}, we present the deviation $\Delta T$ as a function of each parameter, together with the quadratic fit and $95\%$ confidence band. The results demonstrate that the EWCS peak deviation $\Delta T$ is systematically correlated with each model parameter. Additionally, these findings confirm that the EWCS peak trajectory is not an independent or random feature of the holographic entanglement structure but rather a systematic offset from the thermodynamic phase boundary. The monotonic parameter dependence of $\Delta T$ suggests that the higher-order term of EWCS and the effective MIT encode the same underlying physics, with the EWCS exhibiting a shifted but faithful reflection of the phase structure.

\end{document}